\documentclass[%
  article,notitlepage, preprint, amsmath,amssymb, aps, floats,superscriptaddress]{revtex4-1}

\usepackage{graphicx,subeqnarray}
\usepackage{dcolumn}
\usepackage{bm}
\usepackage{color}
\usepackage{amssymb}
\usepackage{amssymb}
\usepackage[normalem]{ulem}

\def\di{\displaystyle}
\def\e{\bf{e}}

\begin{document}

\title{Two-fluid model for  locomotion under self-confinement}

\author{Shang Yik Reigh}

\email{reigh@is.mpg.de}
\affiliation{Department of Applied Mathematics and Theoretical Physics, Centre for Mathematical Science,
  University of Cambridge, Wilberforce Road, Cambridge CB3 0WA, United Kingdom}
\affiliation{Max-Planck-Institut f{\"u}r Intelligente Systeme, Heisenbergstrasse 3, 70569 Stuttgart, Germany}\author{Eric Lauga}
\email{e.lauga@damtp.cam.ac.uk}
\affiliation{Department of Applied Mathematics and Theoretical Physics, Centre for Mathematical Science,
  University of Cambridge, Wilberforce Road, Cambridge CB3 0WA, United Kingdom}

\date{\today}

\begin{abstract}
  The bacterium \textit{Helicobacter pylori} causes  ulcers in the stomach of humans   by invading mucus layers protecting epithelial cells. It does so by chemically changing the rheological properties of the mucus from a high-viscosity gel to a low-viscosity solution {in} which it may self-propel. We develop a two-fluid model for this process of  swimming under self-generated confinement. We solve exactly for the flow  and the locomotion speed of a spherical swimmer located in  a spherically symmetric system of two Newtonian fluids whose boundary moves with the swimmer. We also treat separately  the  special case of an immobile outer fluid. In all cases, we characterise the flow fields, their spatial decay, and the impact of both the viscosity ratio and the degree of confinement on the locomotion speed of the model swimmer. The spatial decay of the flow retains the same power-law decay as for 
  locomotion in a single fluid 
   but with a decreased magnitude. 
Independently of the assumption chosen to characterise the impact of confinement on the actuation applied by the swimmer, its locomotion speed always decreases with an increase in the degree of confinement. Our modelling results suggest that a low-viscosity  region of at  least six times the effective swimmer size is required to lead to swimming with speeds similar to locomotion in an infinite fluid, corresponding to a region of size above $\approx25~\mu$m for \textit{Helicobacter pylori}.

\end{abstract}

\maketitle

\section{Introduction}
Most motile microorganisms swim in viscous fluids by deforming slender appendages 
called flagella and cilia~\cite{brennen77,Lauga2009,goldstein:15}. 
Spermatozoa actively deform  planar flagella as planar waves~\cite{woolley09},
bacteria such as \textit{Escherichia coli} (\textit{E.~coli}) passively rotate rigid helical flagella~\cite{berg04},
while ciliates such as \textit{Paramecium} beat their short cilia with power and recovery strokes to break time-reversal symmetry~\cite{lighthill52,blake71,purcell:77,shay99}. 
These slender appendages  may interact through the fluid, giving rise to synchronised or collective dynamics 
~\cite{taylor51,shay99,woolley09,gole:11,gold09sci,reigh:12}.
Alongside living organisms, self-propelled micro- or nano-machines have been synthesised in the laboratory. Some of them directly mimic the physics of biological microorganisms while others self-propel chemically by  catalysing a chemical reaction in an asymmetric fashion~\cite{dreyfus:05,alois:10,kapral:13,wang:13,wangbook:13,reigh2013synchronization,colberg14,herm:14,walker:15,reigh2015catalytic,reigh2016microscopic,palagi:16}.

Although the detailed propulsion mechanisms of many of these micro- or nano-swimmers are not trivial, 
many of the physical quantifies of interesting including  swimming velocity, fluid flow field, and their collective behaviours may be captured by continuum  fluid mechanics  models with 
 simplified  propulsion mechanisms~\cite{purcell:77,anderson:89,ishikawa06,graham:09,kapral:13,colberg14}. Such swimmers being of small  sizes (tens of microns or less), the  fluid flows generated are in general well described by the Stokes equations~\cite{purcell:77,anderson:89,Lauga2009,colberg14}. The detailed kinematics of the flagella and cilia on the surface of biological swimmers, or the precise chemical kinetics of artificial  swimmers, 
may  then be simplified by imposing prescribed boundary conditions
at the swimmer surface, the swimming velocity being then obtained from the condition that the net force exerted on the swimmer by the fluid is zero~\cite{lighthill52,blake71,anderson:89,Lauga2009,colberg14}.

Theoretically, an isolated swimmer moving in an unbounded fluid of constant viscosity is the ideal model 
to describe self-propulsion of a  self-propelled cell~\cite{lighthill52,blake71,Lauga2009}. In realistic conditions, however, the motion of a swimmer may be  affected by 
various environmental factors  such as external  chemical reactions, physical obstacles and boundaries, viscosity variation in space  or the interactions between  nearby swimmers  in  suspension 
~\cite{ishikawa06,ishikawa:08,graham:09,lowen:09,evans:11,bansil:13,martinez14,holga:14,pop09}.

\begin{figure}[t]
  \centering
  \includegraphics[scale=0.32,angle=0]{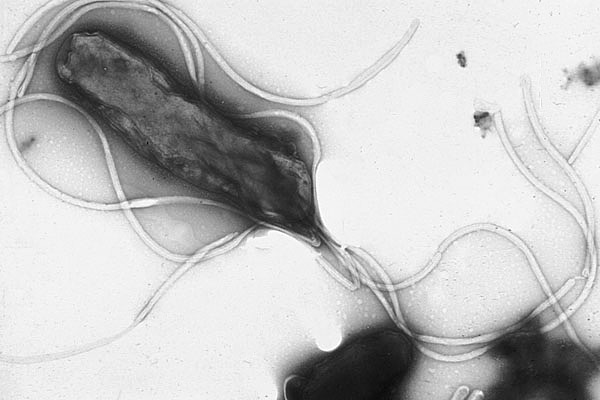}
  \caption{Electron micrograph of  \textit{H.~pylori} possessing multiple flagella. Image courtesy of   Y.~Tsutsumi and reproduced with permission (Wikimedia commons). The length of the cell body is approximately $4~\mu$m \cite{bansil09}. }
  \label{EM}
\end{figure}

The bacterium \textit{Helicobacter pylori} (\textit{H.~pylori}, shown in Fig.~\ref{EM})  is one example of a self-propelled cell moving in a complex  environment~\cite{monte:01,bansil09,bansil:13,mirbagheri16}. This organism is (in)famous for being  able to  survive in the acidic environment of  the human stomach (gastric fluid),  penetrating the mucus layer protecting the stomach lining (a gel-like network of  entangled polymers), reaching the underlying epithelium and eventually causing stomach ulcers. In order to self-propel in the thick mucus protecting the stomach, this bacterium releases the enzyme urease which catalyses the reaction from urea to ammonia and convert the local acidic environment near the bacterium to neutral, thereby  locally destroying the polymer network and transforming it into a viscous fluid in which the bacterium is able to swim~\cite{monte:01}. We note however that the size of this low-viscosity region has never been  measured experimentally. 
Recently, synthetic  micropropellers operated by external magnetic fields have been fabricated by mimicking \textit{H.~pylori}'s physico-chemical swimming strategy~\cite{walker:15}. The micropropellers were  able to change the acidic conditions of the mucus gel to neutral by using the activity of the enzyme urease immobilised on their surfaces.

{From a fluid mechanical standpoint, the intuitive physical picture for \textit{H.~pylori} and its synthetic counterpart  is that of a 
swimmer which, due to the partial or total breakdown of polymer networks, induces  chemically a low-viscosity fluid surrounding it, which  follows it as the swimmer goes along and is itself enclosed in a high-viscosity gel.} This setup of swimming in a domain with spatial variations in  viscosity  has been also observed in other systems. For example, when the bacterium \textit{E.~coli} swims in polymer solutions with high molecular weight, shear thinning and viscosity changes are observed near the rotating flagella~\cite{martinez14}. Experimental studies on the locomotion of the nematode \textit{Caenorhabditis elegans} show a decrease in swimming  in shear thinning fluids~\cite{shen11}, with some debate on the physical origin of this decrease ~\cite{berg:79,liu:11,power:13,gagnon:13,tom:13,dasgupta:13,emily:15,yi15}. Furthermore, in the extreme case where the viscosity of the gel is much larger than the viscosity of the local fluid, one recovers the limit of swimming under  external confinement~\cite{graham:09,pop09,felder10}.

\begin{figure}
\centering
\includegraphics[scale=0.33,angle=0]{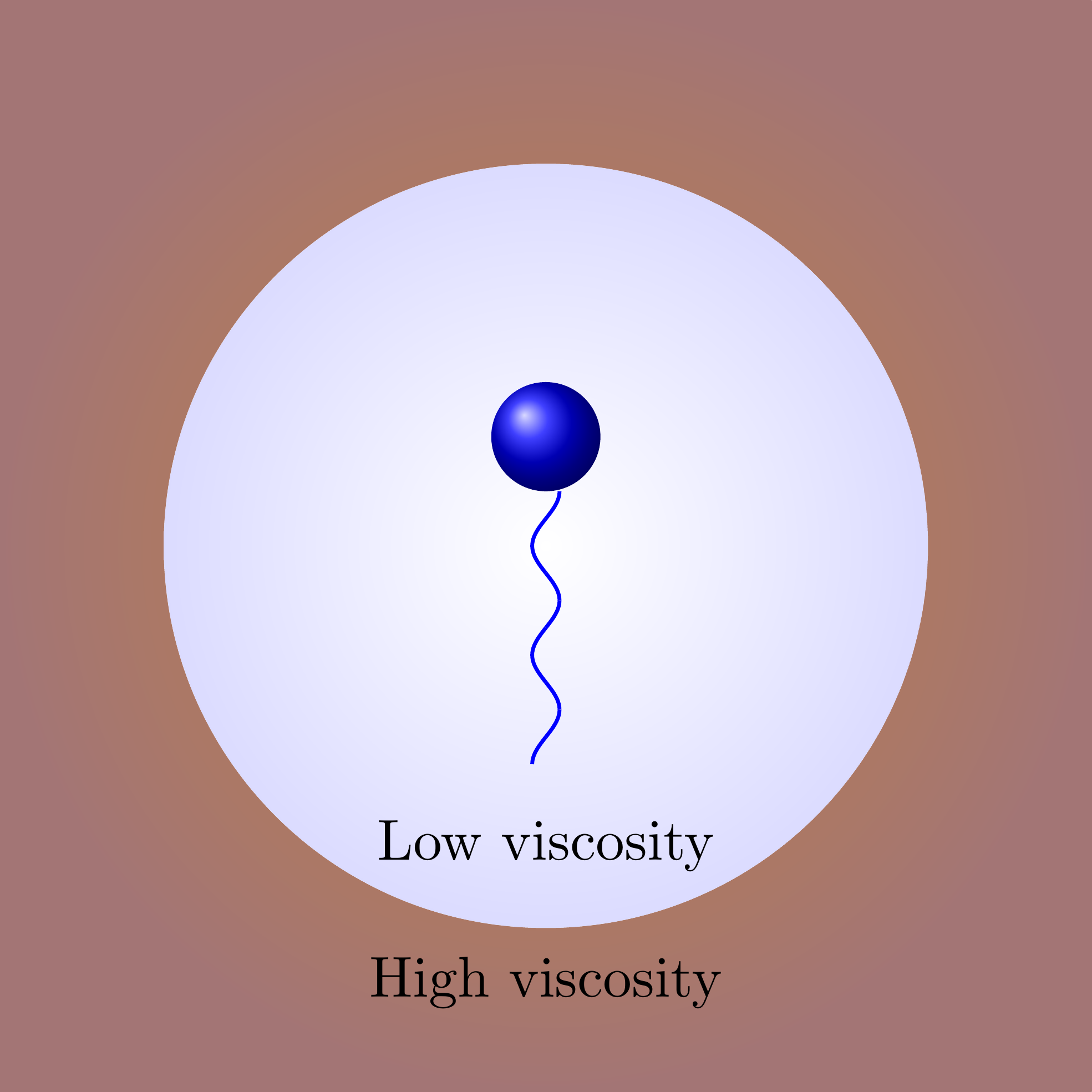}
\caption{A two-fluid model for the locomotion of \textit{H.~pylori}.
The bacterium swims in a low-viscosity inner fluid 
which is surrounded by a high-viscosity outer fluid. The low-viscosity fluid is induced chemically   by  the swimmer and thus continuously  moves along with it. 
  }
  \label{fig1}
\end{figure}

In this paper, we build and study a theoretical fluid-mechanical model for the locomotion of \textit{H.~pylori}. We assume that the swimmer is surrounded by a low-viscosity Newtonian fluid close to the swimmer which is itself enclosed in a second Newtonian   fluid but with a larger viscosity, as illustrated  in Fig.~\ref{fig1}.  As a difference with recent work, we neglect the poro-elastic nature of the gel \cite{mirbagheri16} and focus solely on its viscous behaviour.  The low-viscosity fluid results from the chemical transformation of the second, high-viscosity fluid. Given the small length scales involved in the biological problem (microns), it is a good approximation to assume that the chemical  effect is  purely diffusive.  This can be justified by evaluating the typical value of the P{\'e}clet number, ${\rm Pe} = U L/D$, for the diffusing enzyme. With a typical bacterial swimming speed $U\approx 10 $~$\mu$m/s, a typical organism length scale $L\approx 1$~$\mu$m  and 
{a diffusion constant of ammonia $D\approx 1.6 \times 10^{-9}$~m$^2$/s}, one obtains {${\rm Pe} \approx 6\times10^{-3}$}, consistent with recent modelling work \cite{mirbagheri16} and justifying the small-Pe approximation.  As a result, the size of the low-viscosity region is not affected by the flow created by the swimmer, and is carried along by the swimmer as it self-propels. To model the swimming motion, we consider the classical squirmer model of spherical shape where the swimming results from prescribed tangential and normal velocity  boundary conditions on the surface of the swimmer, and our model can thus be seen as an extension of the squirmer framework to a two-fluid case \cite{lighthill52,blake71}. Within these assumptions, this is a model which can  be completely solved analytically. 
  
The paper is organized as follows. We first derive the exact solution for the Stokes flows in the two fluids and the resulting swimming speed. We then illustrate the  flow fields for relevant swimming motions and discuss their characteristics. As a special case we then consider the limit where the outer fluid is immobile (so infinitely more viscous than the inner fluid). We finally apply our model to the locomotion of \textit{H.~pylori} by considering three different ways in which the confinement may affect the boundary conditions applied by the swimmer. We close the paper by a summary of our work while Appendices list the various constants used in the analytical solution in the main  text of the paper.


\begin{figure}[t]
  \centering
  \includegraphics[scale=0.36,angle=0]{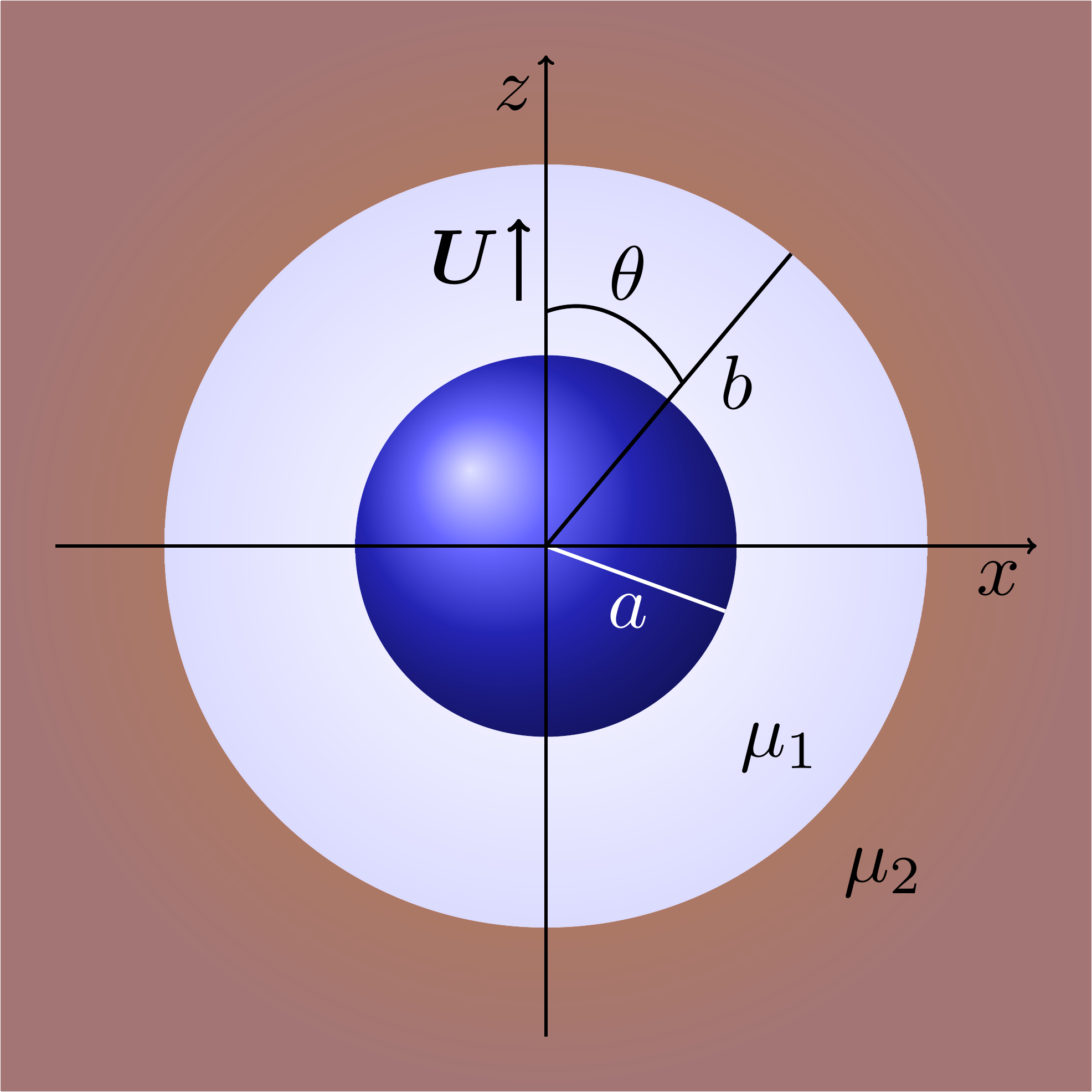}
  \caption{Schematic view of the model in  spherical polar coordinate ($r,\theta,\phi$).
    A spherical model organism of radius $a$ is immersed in a  Newtonian fluid of low viscosity $\mu_1$ itself surrounded by another fluid of viscosity $\mu_2$. The boundary between the two fluids is also spherical and has radius $b \geq a$. 
    The swimmer acts on the fluid by imposing axisymmetric velocity on its surface and as a result swims with velocity $\bm{U} = U\e_z$. Since the boundary between the two fluids is the result of a diffusive problem, it remains spherical  for all times in the frame moving with the swimmer.
  }
  \label{schem}
\end{figure}

\section{\label{twofluid}Two-fluid model}
The specific model we are considering is shown in Fig.~\ref{schem}. The   Newtonian inner fluid has  dynamic viscosity ${\mu}_1$ while the  Newtonian outer fluid has  dynamic viscosity ${\mu}_2$. The swimmer is spherical with radius $a$ and the boundary between the two fluids is also spherical with radius $b \geq a$. Since the presence of the boundary is a result of a purely diffusive chemical process, we can assume that it moves with the  swimmer and thus  remains of radius $b$ in the frame moving with the swimmer. We assume that the forcing on the fluid imposed by the swimmer is axisymmetric and therefore the swimmer moves with velocity  $\bm{U} = U\e_z$, {where $\e_z$ is the unit vector along the $z$ axis.}
We use spherical polar coordinates ($r,\theta,\phi$) to describe the flow fields.

The velocity fields in the two fluids are found by solving the incompressible  Stokes equations
\begin{align}
  \nabla p_i = \mu_i \nabla^2 \bm{v}_i, \quad \nabla \cdot \bm{v}_i=0, \quad (i=1,2)
  \label{stokes}
\end{align}
where $p_i$ is the dynamic pressure in fluid $i$ and $\bm{v}_i$ the fluid velocity.
In the laboratory frame of reference where the swimmer moves with the velocity $U\e_z$,
the fluid velocity $\bm{v}=(v_{r,1},v_{\theta,1})$ on the swimmer surface is taken to be imposed and given by the most general axisymmetric decomposition~\cite{lighthill52,blake71} 
\begin{subeqnarray}  \label{bc1}
  v_{r,1}\vert_{r=a} &=& \sum_{n=0}^{\infty}A_n(t)P_n(\xi) + UP_1(\xi),\\ 
  v_{\theta,1}\vert_{r=a} &=& \sum_{n=1}^{\infty}B_n(t)V_n(\xi) - UV_1(\xi),
\end{subeqnarray}
where the coefficients $A_n(t)$ and $B_n(t)$  are allowed to be time-dependent describing the details of the  propulsion mechanisms. Here
$P_n$ the Legendre polynomial {{of degree $n$}}, $\xi\equiv\cos{\theta}$, {$V_n=-2P_n^1(\xi)/\{n(n+1)\}$}, where $P_n^1$ is the associate Legendre function with order 1.
By choosing  specific functions for the coefficients $A_n(t)$ and $B_n(t)$,
one may model various swimming motions, for example, neutral (irrotational) swimmers such as molecular motors propelled by phoretic mechanisms~\cite{anderson:89,colberg14},
pullers such as the bi-flagellated alga \textit{Chlamydomonas reinhardtii} (\textit{C.~reinhardtii})~\cite{ishikawa06,goldstein:15}, 
pushers such as the flagellated  bacteria \textit{E.~coli} or \textit{H.~pylori}~\cite{drescher11}
as well as time dependent metachronal waving motions in the ciliate \textit{Paramecium}
~\cite{lighthill52,blake71,pak:14}.

At the spherical interface between the inner and outer fluids, we require continuity of both velocities and  stresses as
\begin{subeqnarray}  \label{bc2}
  &v_{r,1}\vert_{r=b} = v_{r,2}\vert_{r=b},\hspace{20pt} 
  v_{\theta,1}\vert_{r=b} = v_{\theta,2}\vert_{r=b},\\
  &\Pi_{rr,1}\vert_{r=b} = \Pi_{rr,2}\vert_{r=b}, \hspace{10pt} 
  \Pi_{r\theta,1}\vert_{r=b}=\Pi_{r\theta,2}\vert_{r=b},
\end{subeqnarray}
where $\Pi_{rr}$ and $\Pi_{r\theta}$ are the components of the stress tensor $\boldsymbol{\Pi}$
in  spherical  coordinates  
and the subscripts $1$ and $2$ denote  the inner and outer fluids respectively. The boundary conditions quantified by Eq.~\eqref{bc2} assume there is no  surface tension at the interface  and the fluid may freely flow  through the interface. The model is thus one where the two fluids are completely miscible, but where the viscosity is different in both domains for all times. {Since the viscosity is set by a purely diffusive problem (small P{\'e}clet number), it is not a material property transported by  the fluid (Lagrangian) but   instead the viscosity is constant in the swimming (Eulerian) frame.}  The  interface indicates thus simply the boundary instantaneously delimiting two constant-viscosity domains which are undisturbed by the fluid motion.

The general  axisymmetric solution of the incompressible Stokes equations  
is given for the velocity field by~\cite{lamb,happel73} 
\begin{align}
  &\bm{v} = \sum_{n=-\infty}^{\infty} \left[\nabla \phi_n 
  + \frac{n+3}{2\mu(n+1)(2n+3)}r^2\nabla p_n -\frac{n}{\mu(n+1)(2n+3)}\bm{r}p_n \right],
  \label{gensol}
\end{align}
where $p_n$ and $\phi_n$ are the solutions of $\nabla^2 p_n = 0$ and $\nabla^2 \phi_n = 0$ respectively and $\mu$ the dynamic viscosity. 
In the spherical polar coordinate, $p_n$ and $\phi_n$ depend on  the Legendre functions as
\begin{align}
  p_n(r,\xi) = \tilde{p}_nr^n P_n(\xi), \quad
  \phi_n(r,\xi) = \tilde{\phi}_nr^n P_n(\xi),
\end{align}
where $\tilde{p}_n$ and $\tilde{\phi}_n$ are constants independent of $r$ and $\xi$. From the above vector form, the radial and azimuthal  velocity 
components   are obtained as
\begin{subeqnarray}
  v_r &=& \sum_{n\ge 0}^{\infty}\left[\bar{p}_nr^{n+1}+\bar{\phi}_nr^{n-1}
  +\bar{p}_{-(n+1)}\frac{1}{r^{n}}  +\bar{\phi}_{-(n+1)}\frac{1}{r^{n+2}}\right] P_n(\xi),\\
  v_{\theta} &=& \sum_{{n\ge 1}}^{\infty} \left[-\frac{n+3}{2}\bar{p}_n r^{n+1}
  -\frac{n+1}{2}\bar{\phi}_n r^{n-1} +\frac{n-2}{2}\bar{p}_{-(n+1)} \frac{1}{r^{n}} 
  + \frac{n}{2}\bar{\phi}_{-(n+1)} \frac{1}{r^{n+2}} \right]V_n(\xi),
\end{subeqnarray}
where
\begin{align}
\bar{p}_n = \frac{n}{2\mu(2n+3)}\tilde{p}_n,
\quad \bar{\phi}_n = n\tilde{\phi}_n.
  \label{unconst}
\end{align}

The unknown constants $\bar{p}_n$ and $\bar{\phi}_n$
are determined by enforcing the boundary conditions, Eqs.~\ref{bc1}-\ref{bc2}.
Taking the coefficients of $P_n(\xi)$ and $V_n(\xi)$ ($n=0,1$) in Eq. 6 
  and comparing them for the inner and outer fluids, we obtain a system of equations for the  unknown constants
\begin{subeqnarray}   \label{arr1}
  \bar{p}_{-1,1} + \frac{1}{a^2}\bar{\phi}_{-1,1} &=& A_0,   \\
  (p_{\infty}-\tilde{p}_0) +(2{\mu}_1+4{\mu}_2)\frac{1}{b}\bar{p}_{-1,1}
  -(4{\mu}_1-4{\mu}_2)\frac{1}{b^3}\bar{\phi}_{-1,1} &=& 0, \\ 
  a^2\bar{p}_{1,1} + \bar{\phi}_{1,1} + \frac{1}{a}\bar{p}_{-2,1} +\frac{1}{a^3}\bar{\phi}_{-2,1} &=& A_1 + U, \\ 
  -2a^2\bar{p}_{1,1} - \bar{\phi}_{1,1} - \frac{1}{2a}\bar{p}_{-2,1} +\frac{1}{2a^3}\bar{\phi}_{-2,1} &=& B_1 - U,\\
  (1-\frac{1}{\sigma})b^2\bar{p}_{1,1} + \bar{\phi}_{1,1} + (1-\frac{1}{\sigma})\frac{1}{b}\bar{p}_{-2,1} 
  +(1-\frac{1}{\sigma})\frac{1}{b^3}\bar{\phi}_{-2,1} &=& 0,\\ 
  (-2-\frac{1}{2\sigma})b^2\bar{p}_{1,1} - \bar{\phi}_{1,1} - (1-\frac{1}{\sigma})\frac{1}{2b}\bar{p}_{-2,1} 
  + (1-\frac{1}{\sigma})\frac{1}{2b^3}\bar{\phi}_{-2,1} &=&0,
 \end{subeqnarray} 
where $p_{\infty}$ is the constant pressure at $r=\infty$,  $\tilde{p}_0$ is the undermined constant pressure in the inner fluid and we use the    subscript $1$ to indicate the inner fluid. 
Since all coefficients {$\bar{p}_{n,1}$ and $\bar{\phi}_{n,1}$}  are zeros in the case of no fluid motion, we can set $p_{\infty}-\tilde{p}_0$ to  zero without loss of generality. We may then use Eq.~\eqref{arr1} to obtain the unknown constants, {$\bar{p}_{n,1}$ and $\bar{\phi}_{n,1}$ ($n=-2,-1,1$)},  in terms of the unknown swimming speed $U$. 
The relevant constants for the outside fluids are obtained from the following relations,
{\begin{subeqnarray}
  \bar{p}_{-2,2} &=& \frac{1}{\sigma}\bar{p}_{-2,1},\\
 \bar{\phi}_{-2,2} &=& \frac{b^3}{\sigma} \Big(b^2\bar{p}_{1,1}+\frac{1}{b^3}\bar{\phi}_{-2,1}\Big), \\
\bar{\phi}_{-1,2} &=& b^2\bar{p}_{-1,1} + \bar{\phi}_{-1,1},
\end{subeqnarray}}where the  subscript $2$  refers to the outer fluid.

The value of the swimming speed, 
 $U$, is  determined by enforcing the force-free condition.
The hydrodynamic force on the swimmer exerted by the fluid is calculated
by integrating the stress $\bm{\Pi}$ on the swimmer surface $S$, 
$\bm{F}=\int\!\!\!\int_S \bm{\Pi}\cdot \hat{\bm{r}}dS$, since for a spherical swimmer $\bm{\Pi} \cdot \hat{\bm{r}}$ is the traction acting from the fluid on the swimmer. This  classically leads to~\cite{happel73}
\begin{align}
  \bm{F} = -4 \pi \nabla(r^3p_{-2}).
  \label{force}
\end{align}
Since no external forces are applied, the constant $\bar{p}_{-2}$ is equal to zero. 
From this condition, the swimming velocity is easily obtained as 
\begin{align}
  U = \frac{\Xi_1\sigma + \Xi_2}{\Delta_1},
  \label{vel_sq}
\end{align}
where
\begin{subeqnarray}
  \Xi_1(\lambda)  &=& 2(2B_1-A_1)\lambda^5 -10(A_1+B_1)\lambda^2 +6(2A_1+B_1),\\
  \Xi_2(\lambda)  &=& 3(2B_1-A_1)\lambda^5 +10(A_1+B_1)\lambda^2 -6(2A_1+B_1),\\
  \Delta_1(\lambda,\sigma)  &=& {3\{2(\lambda^5-1)\sigma+3\lambda^5+2\}},
\end{subeqnarray}
and where we have introduced the two relevant dimensionless numbers in our problem: 
$\lambda=b/a$ is the size ratio between the inner fluid domain  and the swimmer  
and $\sigma=\mu_2/\mu_1$ is the ratio of viscosity between the outer and inner fluid.
Note that, similar  to  the infinite-fluid calculation, the swimming velocity only depends on the constants $A_1$ and $B_1$~\cite{lighthill52,blake71,pak:14}.

The values of the coefficients for  $n \geq 2$ are obtained in a similar way.
In the laboratory frame of reference, the  velocity field for the inner fluid
is then written by 
\begin{subeqnarray}\label{vf_1f}
  v_{r,1} & = & \frac{2A_0}{\Delta_0}\Big\{\Big(\sigma+\frac{1}{2}\Big)\lambda^2\Big(\frac{a}{r}\Big)^2
  -(\sigma-1)\Big\}P_0(\xi)\nonumber\\
  &  & +\frac{(A_1+B_1)}{\Delta_1}\Big\{6(\sigma-1)\Big(\frac{r}{a}\Big)^2-10(\sigma-1)\lambda^2
  +2(2\sigma+3)\lambda^5\Big(\frac{a}{r}\Big)^3\Big\}P_1(\xi) \nonumber\\
  &&+\sum_{n=2}^{\infty} \Big\{\frac{N_1A_n+N_2B_n}{\Delta_{n,1}}\Big(\frac{r}{a}\Big)^{n+1}
  +\frac{N_3A_n+N_4B_n}{\Delta_{n,1}}\Big(\frac{r}{a}\Big)^{n-1}
  +\frac{N_5A_n+N_6B_n}{\Delta_{n,2}}\Big(\frac{a}{r}\Big)^{n} \nonumber\\
  &&
  \quad\quad\quad+\frac{N_7A_n+N_8B_n}{\Delta_{n,2}}\Big(\frac{a}{r}\Big)^{n+2}\Big\} P_n(\xi),\\
  v_{\theta,1} & = & \frac{(A_1+B_1)}{\Delta_1}\Big\{-12(\sigma-1)\Big(\frac{r}{a}\Big)^2+10(\sigma-1)\lambda^2
  +(2\sigma+3)\lambda^5\Big(\frac{a}{r}\Big)^3\Big\}V_1(\xi) \nonumber\\
  &&+\sum_{n=2}^{\infty}\Big\{-\frac{n+3}{2}\frac{N_1A_n+N_2B_n}{\Delta_{n,1}}\Big(\frac{r}{a}\Big)^{n+1}
  -\frac{n+1}{2}\frac{N_3A_n+N_4B_n}{\Delta_{n,1}}\Big(\frac{r}{a}\Big)^{n-1} \nonumber\\
  &&\quad\quad\quad+\frac{n-2}{2}\frac{N_5A_n+N_6B_n}{\Delta_{n,2}}\Big(\frac{a}{r}\Big)^{n} \nonumber\\
  &&\quad\quad\quad +\frac{n}{2}\frac{N_7A_n+N_8B_n}{\Delta_{n,2}}\Big(\frac{a}{r}\Big)^{n+2}\Big\}
  V_n(\xi),
\end{subeqnarray}
where the values of the  undefined constants are given in Appendix~\ref{table1}. Similarly, we obtain for the outer fluid
\begin{subeqnarray}
  v_{r,2} &=& \frac{3\lambda^2A_0}{\Delta_0}\Big(\frac{a}{r}\Big)^2P_0(\xi)
  +\frac{10\lambda^5(A_1+B_1)}{\Delta_1}\Big(\frac{a}{r}\Big)^3P_1(\xi)\nonumber\\&&
  +\sum_{n=2}^{\infty}\Big\{(c_1A_n+c_2B_n)\Big(\frac{a}{r}\Big)^{n}+(c_3A_n+c_4B_n)\Big(\frac{a}{r}\Big)^{n+2}\Big\}P_n(\xi),\\
  v_{\theta,2} &=&\frac{5\lambda^5(A_1+B_1)}{\Delta_1}\Big(\frac{a}{r}\Big)^3V_1(\xi)\nonumber\\
  &&
  +\sum_{n=2}^{\infty}\Big\{\frac{n-2}{2}(c_1A_n+c_2B_n)\Big(\frac{a}{r}\Big)^{n}
  +\frac{n}{2}(c_3A_n+c_4B_n)\Big(\frac{a}{r}\Big)^{n+2}\Big\}V_n(\xi).
  \label{vf_2f}
\end{subeqnarray}
Formally, one sees that the  swimming velocity (Eq.~\ref{vel_sq}) and the fluid velocity fields (Eqs.~\ref{vf_1f}-\ref{vf_2f})
reduce to those of the unbounded flow (Lighthill and Blake's solutions) when we take the viscosities in the fluid to be equal, i.e.~$\sigma=1$~\cite{lighthill52,blake71}.
Furthermore, in the limit where the outer fluid is much more viscous than the inner one, $\sigma \rightarrow \infty$, these solutions match the limit of a no-slip outer surface studied in detail in  Sec.~\ref{sec_im}.


\section{Swimming speeds and fluid flow fields}
\subsection{Swimming speeds}
\begin{figure}[t]
  \centering
  \includegraphics[scale=0.5,angle=0]{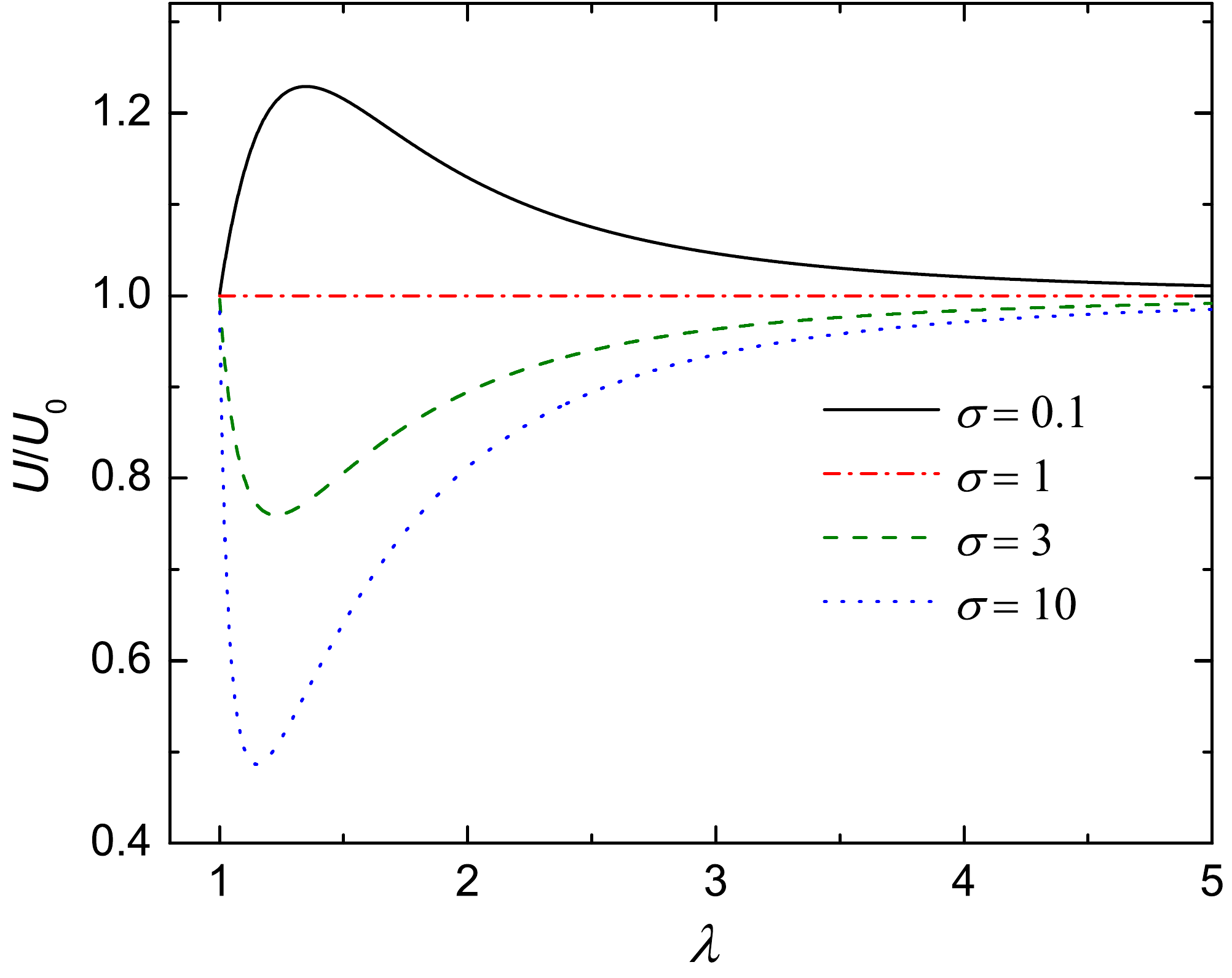}
  \caption{Swimming velocity as a function of  confinement, $\lambda=b/a$, 
    for four values of the viscosity ratio, $\sigma=\mu_2/\mu_1$, in the case where only the tangential modes  $B_n$ are taken into account ($A_n=0$). 
    The black, red, green, and blue  lines (top to bottom) 
    correspond to the viscosity ratios
    $\sigma$ = 0.1, 1, 3 and 10. The swimming velocity is scaled by the  one-fluid value,  
 {$U_0=2 B_1/3$}. 
}
  \label{vel_lam}
\end{figure}

The first quantity of interest for the swimmer is the value of its locomotion speed. In Fig.~\ref{vel_lam} we plot the value of the swimming speed as a function of the confinement size, $\lambda=b/a\geq 1$, 
for four values of the viscosity ratio $\sigma=\mu_2/\mu_1$. Here the influence of the  radial ($A_n$) mode is neglected and only the tangential ($B_n$) modes are considered. When the viscosity in the inner and outer fluids are equal (i.e.~when $\sigma=1$), the swimming velocity does not depend on the confinement size. Indeed in that case, both fluids are in fact just  a single fluid with a common viscosity and the locomotion of the swimmer is thus indistinguishable from that in an unbounded flow, with velocity 
equal to  the classical value  {$U_0=2 B_1/3$}~\cite{lighthill52}; we use  this value to nondimensionalise $U$ in Fig.~\ref{vel_lam}.

When the viscosity in the inner fluid is smaller than that in the outer fluid
($\mu_2>\mu_1$, i.e.~$\sigma > 1$), which is the situation relevant to the
locomotion of {\it H.~pylori}, we see that the swimming speed is always
smaller than that occurring in a single fluid.  {For example, for $\sigma=10$, the maximum reduction in swimming (of about 50\%)  occurs for a confinement ratio of  $\lambda \approx 1.1$}, with an increase back to the one-fluid limit for $\lambda \to 1$ and $\lambda \to \infty$. Notably, if we were to consider the opposite situation where the fluid near the swimmer is of higher viscosity than the outer fluid ($\mu_2<\mu_1$, i.e.~$\sigma < 1$), we see from Fig.~\ref{vel_lam} that the situation is reversed: the swimming speed is always enhanced by the presence of a second fluid.

\begin{figure}
  \centering
  \includegraphics[scale=0.5,angle=0]{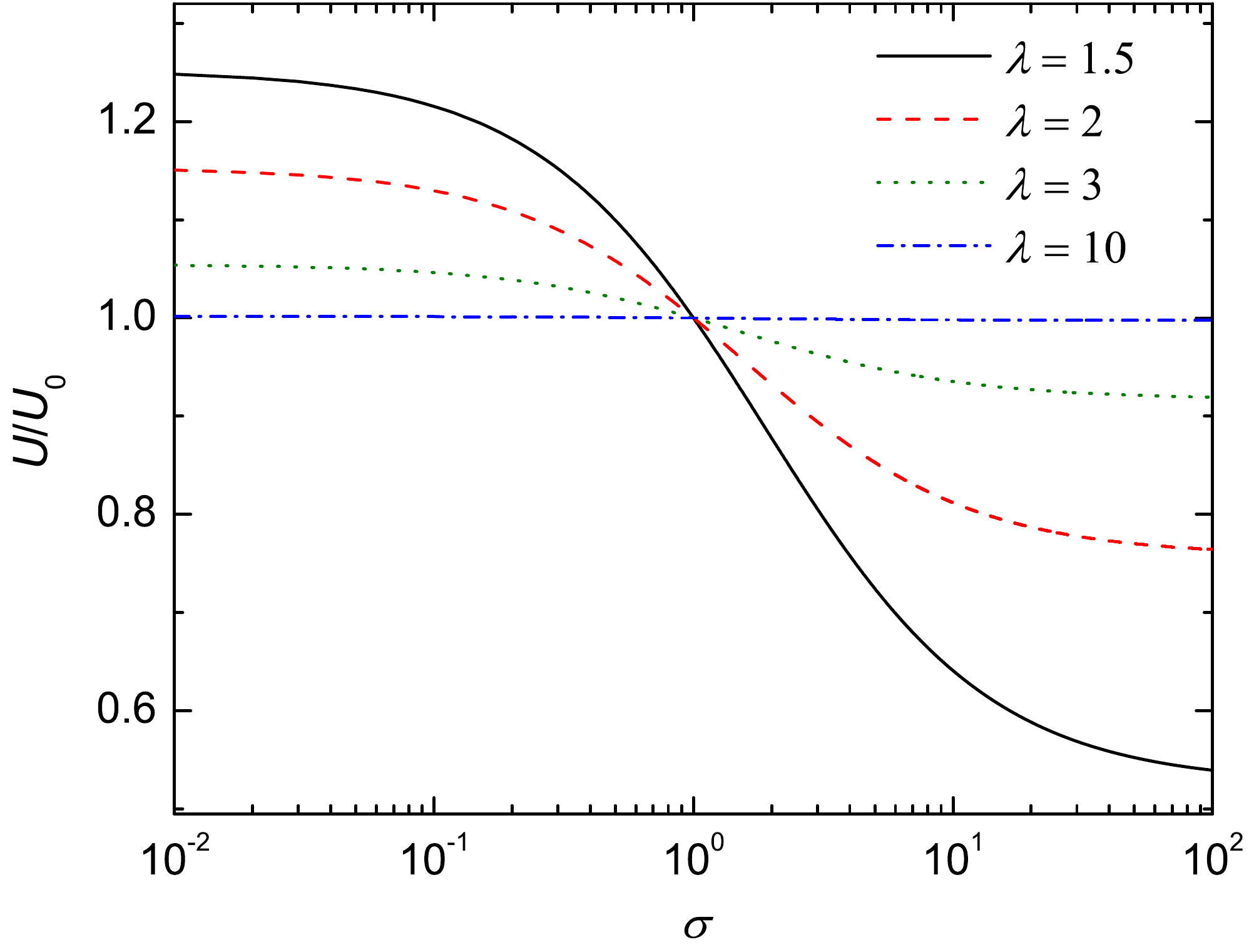}
  \caption{Swimming velocity, $U$, scaled by the one-fluid value, $U_0=2 B_1/3$,
  as a function of the viscosity ratio, $\sigma=\mu_2/\mu_1$, 
    for four different confinements, $\lambda=b/a$. Here again $A_n=0$ and only tangential modes are kept.
    The black, red, green, blue lines  correspond to the size ratios
    $\lambda$ = 1.5, 2, 3 and 10.}
  \label{vel_sig}
\end{figure}

In Fig.~\ref{vel_sig} we plot the swimming velocity as a function of the viscosity ratio, $\sigma$, 
for fixed confinement sizes. We clearly observe the difference in increase vs.~decrease of the swimming {speed} before and after the $\sigma=1$ threshold.  When the boundary between the two fluids is far from the swimmer ($\lambda\gtrsim10$), it does not feel the confinement and swims essentially  as in a single fluid. For small, order-one, size ratios, we see a significant impact of the presence of a second-fluid on the swimming speed.

In Figs.~\ref{vel_lam} and \ref{vel_sig}, we note that the swimming speed was scaled by 
  the speed in an unbounded fluid ($U_0$), which allowed to capture  the  effect of confinement assuming that the swimmer has a constant surface velocity (i.e.~a fixed value of $B_1$)
  which is not impacted by the degree of confinement and variations in viscosity. In order to capture  the impact of size ratios on bare swimming speeds, we consider the biologically-relevant case of a swimmer imposing a constant   mean traction on its surface   instead of a constant surface velocity  (see related discussion in Sec.~\ref{bio}). 
  By defining the mean root-mean-square   traction as
  \begin{equation}
    M\equiv \frac{1}{\gamma}\int\!\!\!\int_S \sqrt{\Pi_{rr}^2+\Pi_{r\theta}^2} dS,
  \end{equation}
  where $\gamma=\int_{-1}^1 \sqrt{1+3\xi^2}d\xi$ and $S$ stands for the
  surface of the swimmer, one gets analytically
  \begin{equation}
    M=\frac{12\pi a \mu_1 B_1}{\Delta_1}\left((2\lambda^5+3)\sigma +
    3(\lambda^5-1)\right).
    \label{mean_force}
  \end{equation}
  Considering swimming with only the $n=1$ mode   as above, we may use  Eqs.~\eqref{vel_sq} and \eqref{mean_force} to
  obtain the swimming speed (normalised by $M$ and the innner viscosity) as a function of the degree of confinement and viscosity ratio as
  \begin{equation}
    \frac{6\pi a \mu_1 U}{M}= \frac{(2\lambda^5-5\lambda^2+3)\sigma+3\lambda^5+5\lambda^2-3}{(2\lambda^5+3)\sigma+3(\lambda^5-1)}\cdot
    \label{vel_lam2}
  \end{equation}

\begin{figure}[t]
  \centering
  \includegraphics[scale=1.1,angle=0]{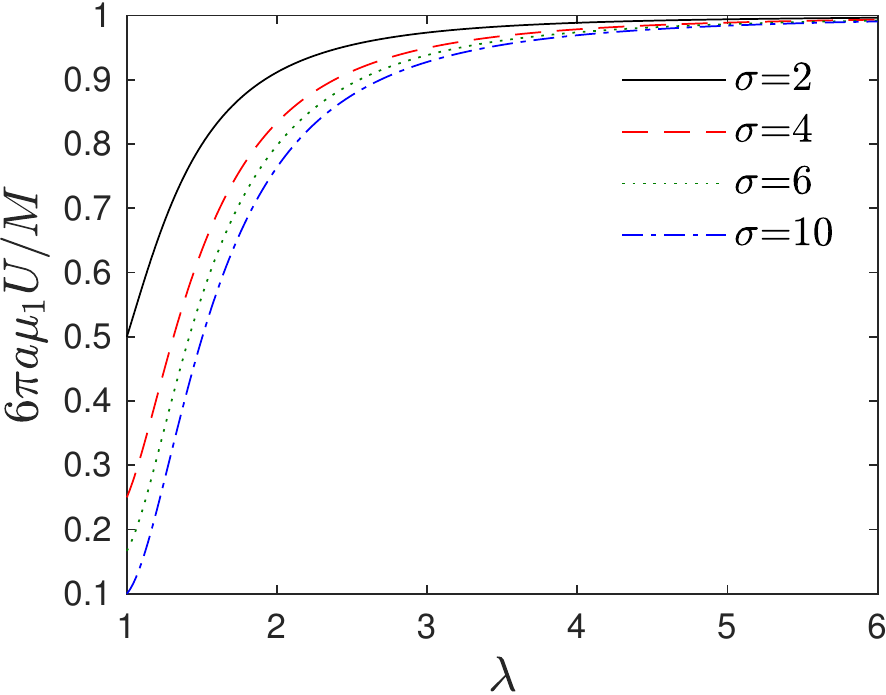}
  \caption{Scaled swimming velocity, $U$, for fixed mean surface traction, $M$, as a function of confinement, $\lambda=b/a$, for various values of the viscosity ratio, $\sigma$.       Only the tangential modes $B_1$ is taken into account, while all other modes are set to zero.       The solid black,  dashed red, dotted green, and dash-dotted blue  lines  
show the results for  viscosity ratios given respectively by $\sigma$ = 2, 4, 6 and 10 (top to bottom).     }
  \label{vel_lam_mstr}
\end{figure}

We plot in Fig.~\ref{vel_lam_mstr}  the swimming speed   as a function of the size ratios as predicted by this fixed-mean-traction result (Eq.~\ref{vel_lam2}). As the size of the low-viscosity region  increases around the swimmer, the swimming speed of the cell always increases. We observe that  a plateau is  reached when the size of the outer region is approximately six times the size of the bacterium (i.e.~$\lambda \approx 6$). 
We note in addition that the swimmer experiences a larger  speed increase as the viscosity ratio between the inner
  and outer regions increases. This can be explained by examining Eq.~\eqref{vel_lam2}. In the limit $\lambda \approx 1$ we get a scaled velocity given by $1/\sigma$ whereas it is given by $1$ in the unconfined limit $\lambda \gg 1$, hence explaining a ratio of velocity equal to a ratio of viscosity.


\subsection{Flow fields}
The  flow fields generated by the swimmers may be visualised by choosing specific boundary conditions on the swimmer surface.  We choose here three representative examples with only tangential velocity components  and no radial deformation ($A_n=0$)  for illustrations as follows:
\begin{equation}
  \begin{aligned}
    &\text{Neutral:} \ B_1=1, B_n=0 \ (n\ne 1),\\
    &\text{Pusher:} \ B_1=1, B_n=-1 \ (2\le n \le 5),\\
    &\text{Puller:} \ B_n=1 \ (1\le n \le 5),
\end{aligned}
\end{equation}
where the three names are chosen in relation to the nature of the flow in the far field of the swimmer.  Neutral swimmers are appropriate models to capture synthetic  locomotion based on phoretic mechanisms such as
diffusiophoresis and thermophoresis~\cite{anderson:89,bickel:13,colberg14} which in {some instances (including so-called Janus particles)} do not induce traditional stresslet flows in the far field.  Pushers represent the majority of flagellated bacteria 
such as \textit{E.~coli} or \textit{H.~pylori} where cells are pushed forward by actuation mechanisms (typically, flagella) located in the rear part of the swimmers~\cite{drescher11}. They induce a stresslet flow in the far field.  In contrast pullers are pulled forward by actuation mechanisms located in the front part of the swimmer, such as   the green alga \textit{C.~reinhardtii}~\cite{ishikawa06,goldstein:15}. The stresslet in this case is of sign opposite to the one induced by pusher cells. 

\begin{figure}
  \centering
  \includegraphics[height=0.4\textwidth]{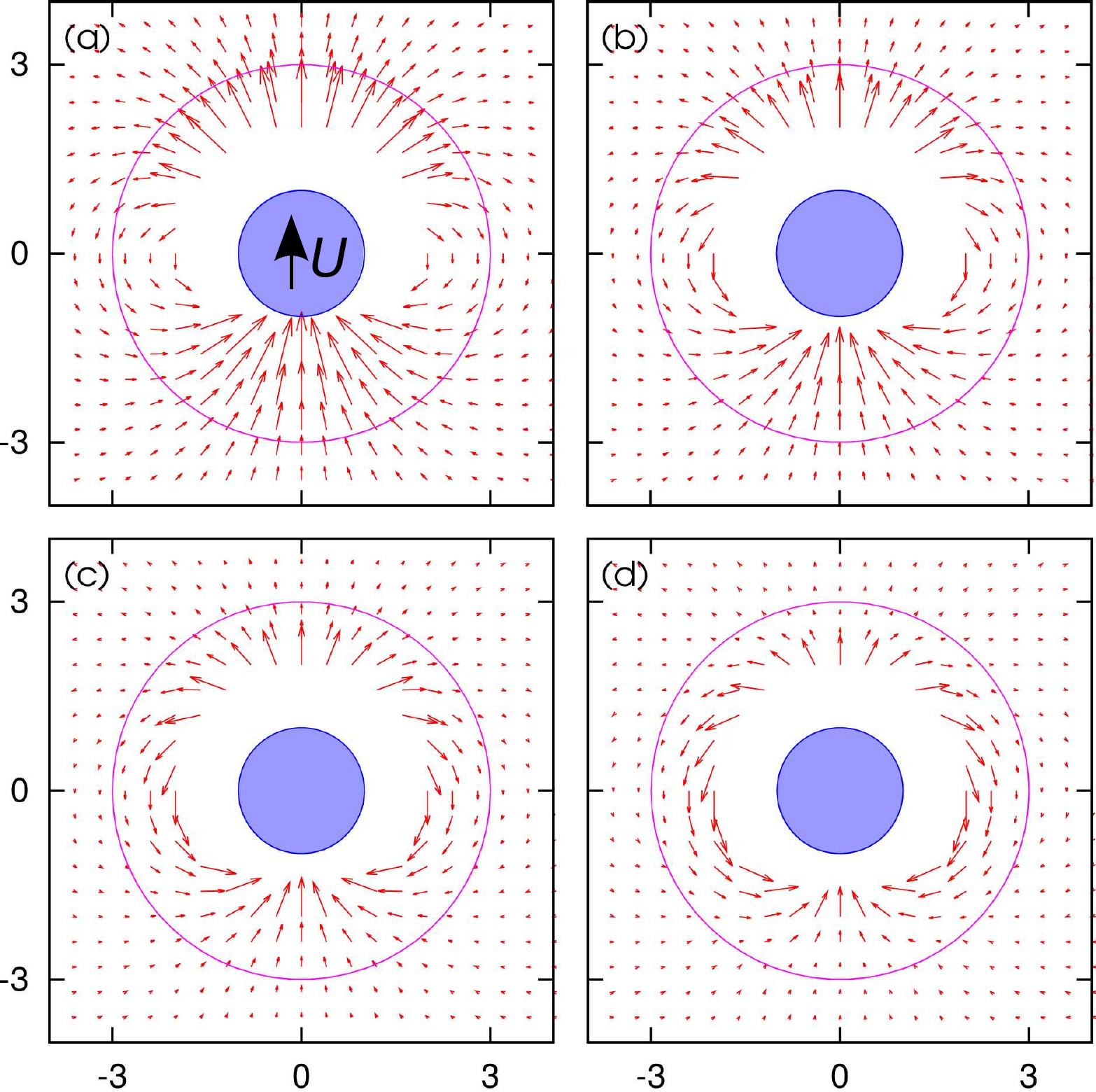}
    \includegraphics[height=0.41\textwidth]{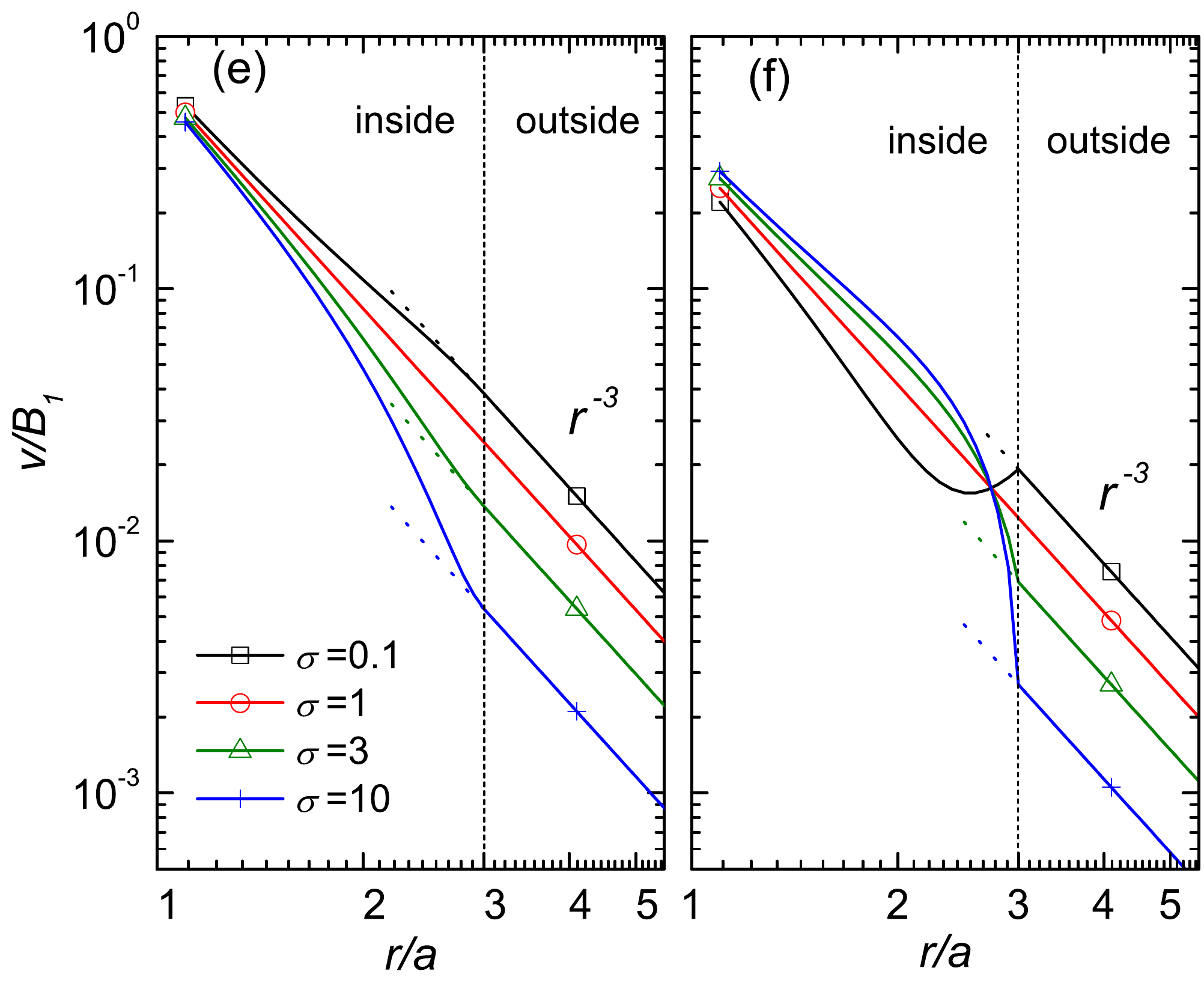}
  \caption{Left: Flow fields for the neutral swimmer in the laboratory frame of reference
with $b/a=3$ and    for various viscosity ratios: 
    (a) $\sigma=0.1$; 
    (b) $\sigma=1$;  
    (c) $\sigma=3$;  
    (d) $\sigma=10$,
    where $\sigma=\mu_2/\mu_1$.   
    The velocity fields near the swimmers are not shown to allow clear visualization.
    Right: Plots of the magnitude of the total fluid velocity, $v=\sqrt{v_r^2+v_\theta^2}$, 
    for the same neutral swimmer
    in (e) the forward directions ($\theta=0$) and (f) the side directions ($\theta=\pi/2$)
    in the laboratory frame of reference.
    The black, red, green, blue solid lines correspond to the viscosity ratio
    $\sigma=$ 0.1, 1, 3, 10, respectively.
    The dotted lines are the respective asymptotics.
    The vertical short dotted lines indicate the boundary of the inner fluid and the outer fluid.}
  \label{fig_md1}
    \label{asym_md1}
\end{figure}

\subsubsection{Neutral swimmers}
In Fig.~\ref{fig_md1} (left) we illustrate the velocity fields,  in the laboratory frame of reference,  for the neutral swimmers for various viscosity ratios.
When the viscosities in the inner and outer fluids are equal ($\sigma=1$, Fig.~\ref{fig_md1}b), 
the flow field is exactly the same as in a single fluid. 
As the viscosity in the outer fluid increases (Fig.~\ref{fig_md1}c and d),
the mobility of the outer fluid decreases 
and the flow in the inner fluid {tends to recirculate} in the inside, until 
 the viscosity in the outer fluids becomes so large that   the flow in the inner fluid can no longer go through the interface. In contrast, when the viscosity in the outer fluid is smaller than the inner fluid
(Fig.~\ref{fig_md1}a), the typical fluid velocities are stronger than in the case of  equal viscosities (Fig.~\ref{fig_md1}b). We recall that the swimming velocity increases in this case (Fig.~\ref{vel_lam}).

The magnitudes of the  fluid velocities as a function of {the distance from the swimmer} are  shown in Fig.~\ref{asym_md1} (right) for different  viscosity ratios. The  {flow} velocity ahead of the  swimmer ($\theta=0$ in spherical coordinates)  is shown in Fig.~\ref{asym_md1}e  while the velocity on the side of the swimmer ($\theta=\pi/2$) is displayed in Fig.~\ref{asym_md1}f. Ahead of the swimmer, the magnitude of the velocity in both the inner and outer fluid decreases with an increase of the viscosity ratio. In contrast, on the side of the swimmer,  a reversed region appears in the inner fluid where an increase of the viscosity ratio accentuates the circulation in the inner fluid.  In the far field, the fluid velocity  exhibits a {$r^{-3}$} decay behavior  characteristic of neutral  swimmers for all  viscosity ratios, with a magnitude which is reduced by an increase of the viscosity ratio.  {Note that at the interface between the two fluids,  the flows velocities are continuous but not the flow gradients. This is a simple consequence of having two different viscosities on either side of the interface.}
\begin{figure}[t]
  \centering
  \includegraphics[height=0.4\textwidth]{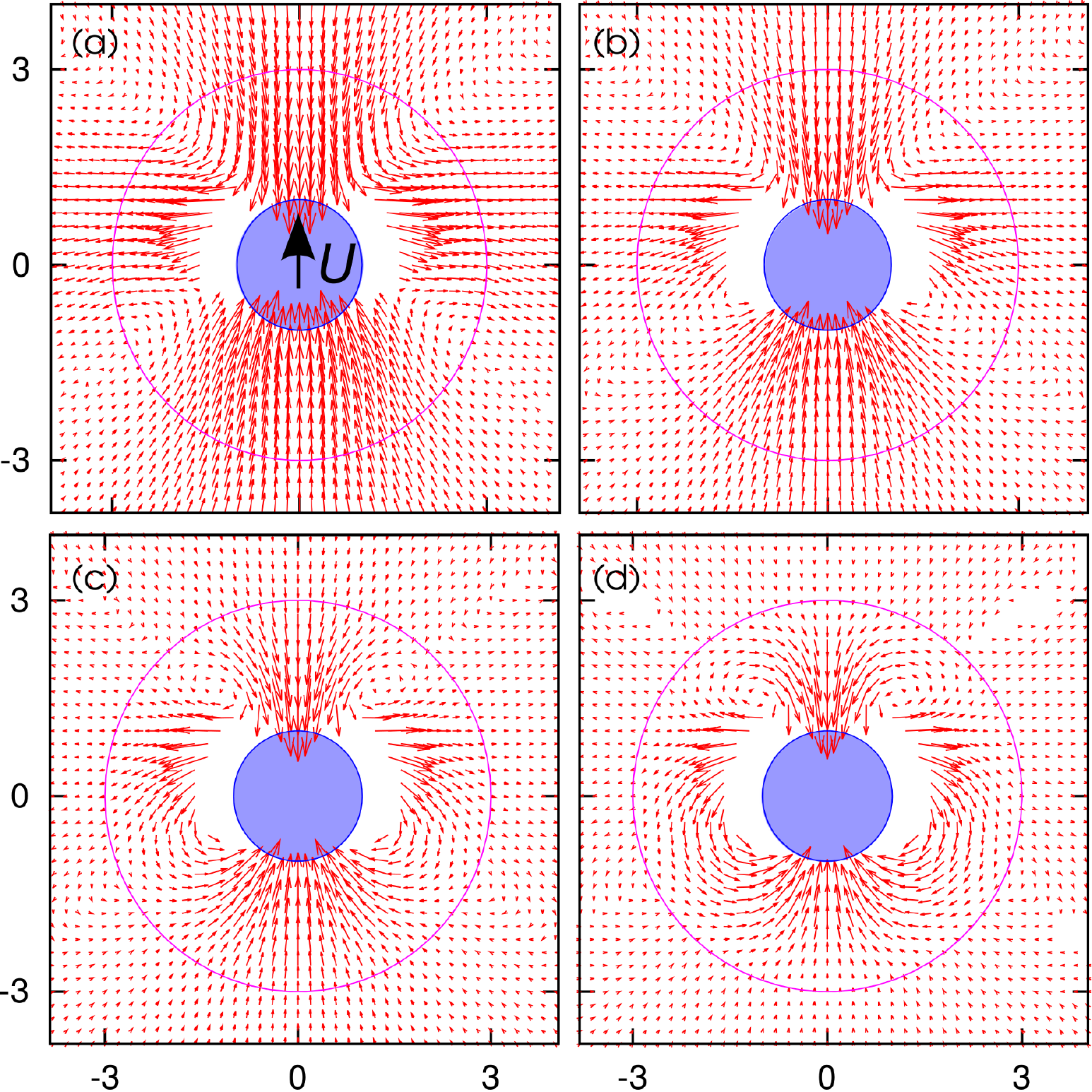}
    \includegraphics[height=0.41\textwidth]{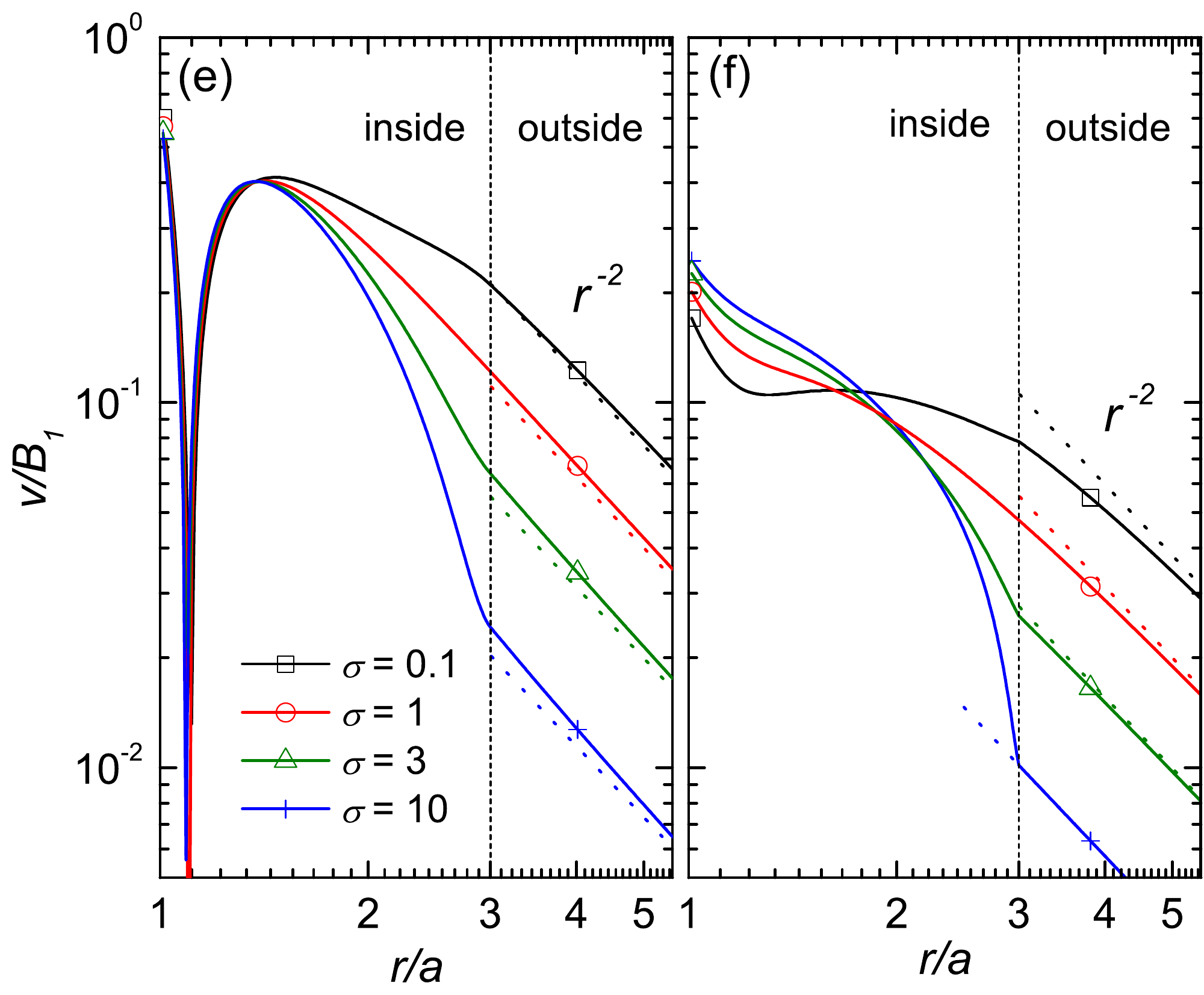}
    \caption{Left: Velocity fields in both fluids for the puller swimmer in the laboratory frame of reference
    for a size ratio $\lambda=3$ and various viscosity ratios $\sigma=\mu_2/\mu_1$: 
    (a) $\sigma=0.1$; 
    (b) $\sigma=1$;  (c) $\sigma=3$;  
    (d) $\sigma=10$. 
    The velocity fields near the swimmers are not shown to allow clear visualization.
    Right: Magnitude of the fluid velocity, $v=\sqrt{v_r^2+v_\theta^2}$,  
    for the same puller swimmer
    in (e) the forward directions ($\theta=0$) and (f) the side directions ($\theta=\pi/2$).
    The black, red, green, blue solid lines correspond to the viscosity ratio
    $\sigma=$ 0.1, 1, 3, 10, respectively.
    The dotted lines are the respective asymptotics while the vertical short dotted lines indicate the interface between the two fluids.
    }
  \label{fig_pull}
    \label{asym_pull}
\end{figure}

Interestingly, {from} an inspection of the  analytical solution for the flow in the outer fluid (Eq.~\ref{vf_2f})  one sees that  the flow components are given by
 \begin{subeqnarray}
v_{r,2} &=& {U_B}\lambda^3\left(\frac{a}{r}\right)^3\cos\theta,\\
v_{\theta,2} &=& \frac{1}{2}{U_B}\lambda^3\left(\frac{a}{r}\right)^3\sin\theta,
\end{subeqnarray}
where ${U_B}= 10(A_1+B_1)\lambda^2/\Delta_1$.
In a vector form, it may be simply expressed as
\begin{equation}
\bm{v}=\frac{U_B}{2}\left(\frac{b}{r}\right)^3(3\hat{\bm{r}}\hat{\bm{r}}-\bm{I})\cdot {\e_z},
\end{equation}
where $\hat{\bm{r}}$ is the unit radial vector and $\bm{I}$ is the unit dyadic.
This velocity expression is that of a self-propelled neutral swimmer  with  radius $b$ swimming with  velocity {$U_B$}.  From the point of view of the outer fluid, the combination of swimmer (size $a$) + inner fluid  is equivalent to a larger force-free swimmer (size $b$) self-propelling at a different speed whose magnitude is a function of  the viscosity ratio and decays as $1/\sigma$.

\subsubsection{Swimmers with stresslets in the far field}

\begin{figure}[t]
  \centering
  \includegraphics[height=0.4\textwidth]{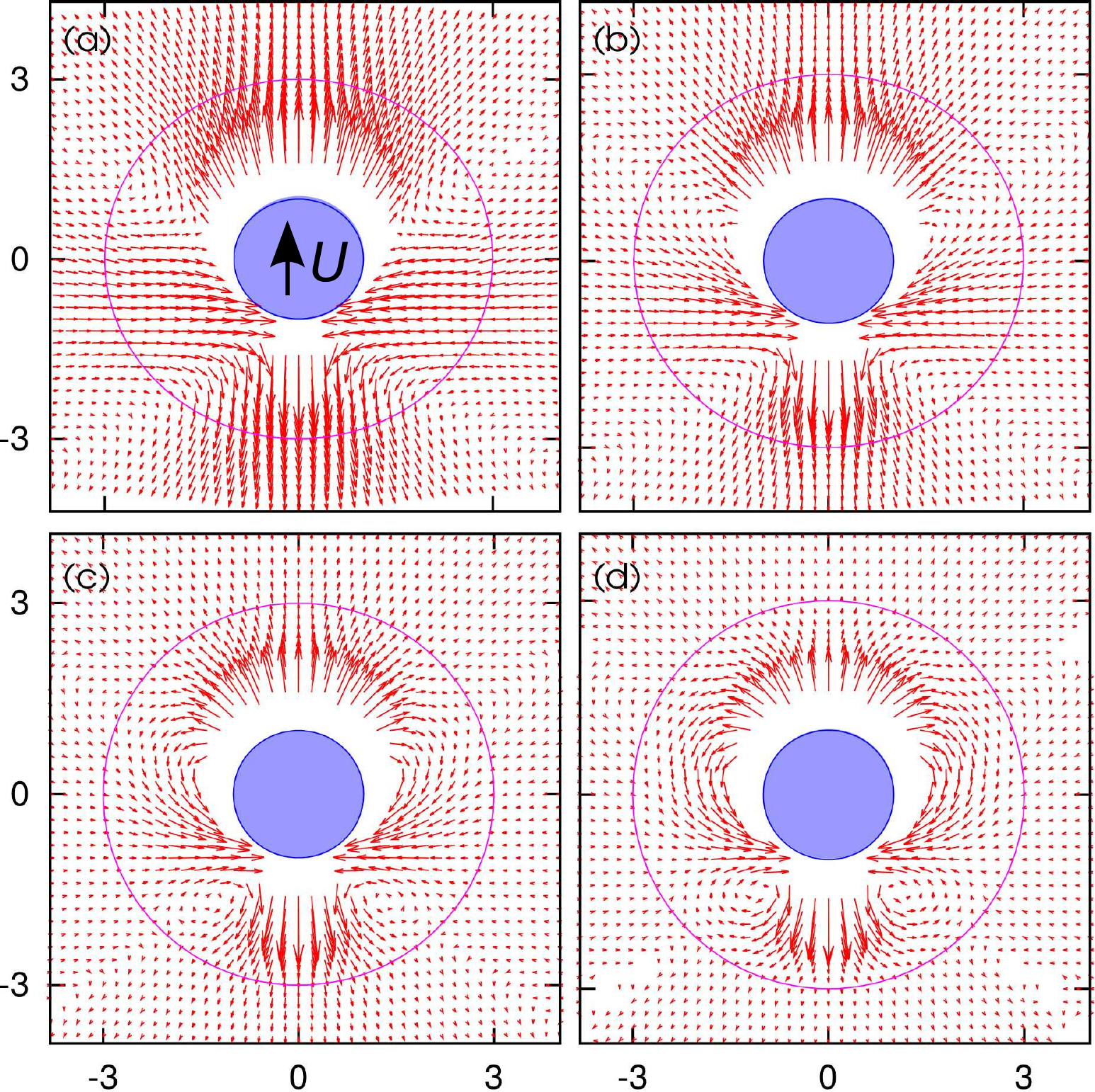}
    \includegraphics[height=0.41\textwidth]{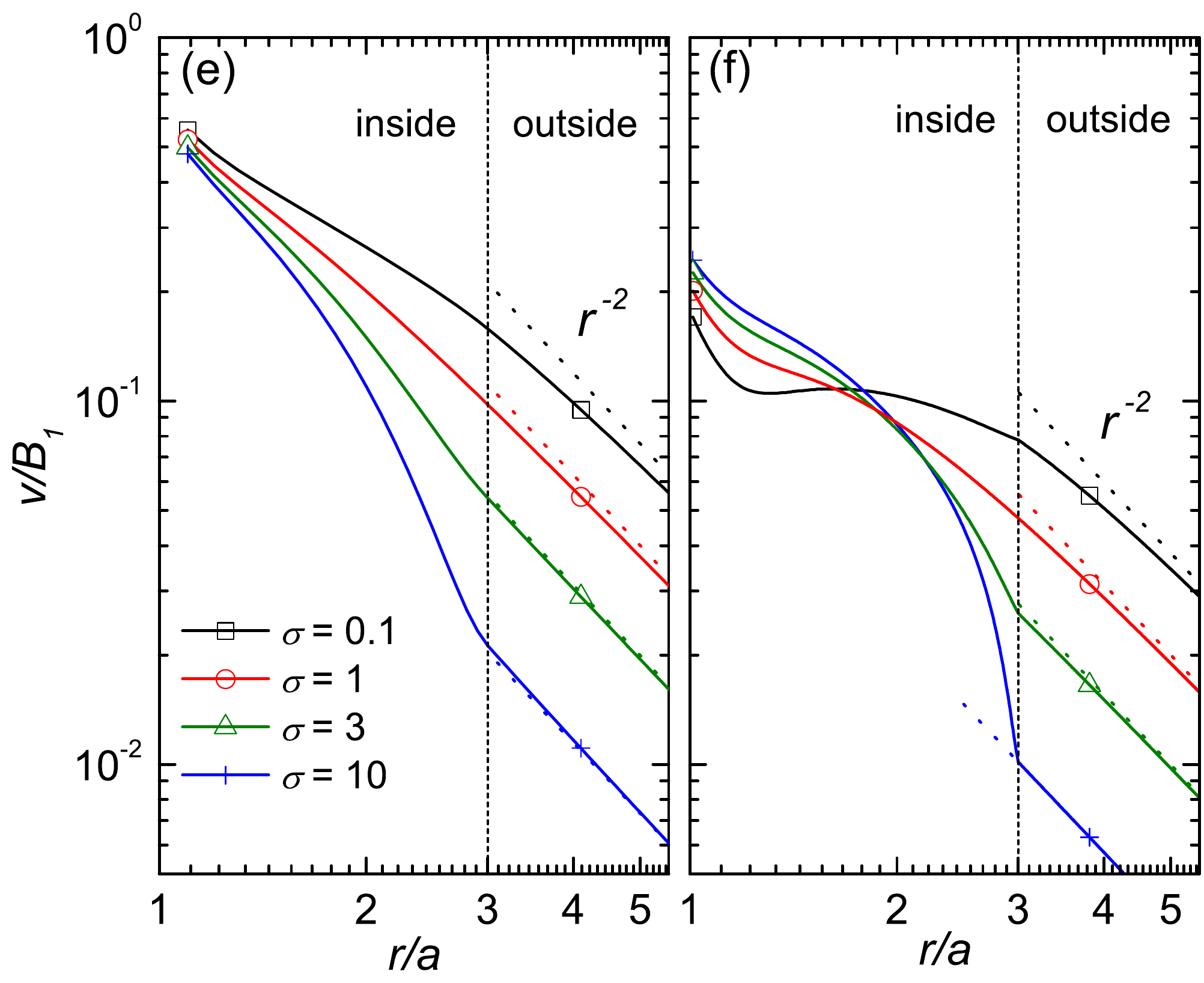}
  \caption{Fluid velocity fields for the pusher swimmer in the laboratory frame of reference
    for $\lambda=3$ and various viscosity ratios:
    (a) $\sigma=0.1$; 
    (b) $\sigma=1$;  (c) $\sigma=3$;  
    (d) $\sigma=10$,
    where $\sigma=\mu_2/\mu_1$. 
    The velocity fields near the swimmers are not shown to allow clear visualization.
    Right: Magnitude of the fluid velocity, $v=\sqrt{v_r^2+v_\theta^2}$,  
    for the same pusher swimmer
    in (e) the forward directions ($\theta=0$) and (f) the side directions ($\theta=\pi/2$).
    The black, red, green, blue solid lines correspond to the viscosity ratio
    $\sigma=$ 0.1, 1, 3, 10, respectively.
    The dotted lines are the respective asymptotics.
    The vertical short dotted lines indicate the boundary between the two fluids. 
}
  \label{fig_push}
  \label{asym_push}
\end{figure}

Moving on to swimmers which induce a stresslet in the far field, we show in Fig.~\ref{fig_pull} (left) the flow fields created in both fluids  {by} the puller swimmer in the laboratory frame of reference. The flow  is directed toward the swimmer from its fore and rear parts   and  away from it on the  sides. The overall dependence of the  flow fields on the viscosity ratio is  similar to the neutral swimmer.  When the viscosity ratio is one ($\sigma=1$, Fig.~\ref{fig_pull}b),  the flow field is exactly the same as in a single fluid. As the viscosity in the outer fluid increases 
($\sigma=3, 10$, Fig.~\ref{fig_pull}c and d),  the flow becomes increasingly  confined in the inner fluid with an increase of the circulation there. Note that a toroidal vortex appears ahead of the swimmer which  becomes stronger as the viscosity in the outer fluid increases.  The quantitative values of the fluid velocity ahead  ($\theta=0$) and on the side of the swimmer ($\theta=\pi/2$)   are shown in Fig.~\ref{asym_pull}e and f respectively. The puller swimmer retains the signature of a puller in the far field and its $r^{-2}$ power law decay.  The  magnitudes of the velocity generally decrease with an increase of the viscosity in the outer fluid  except in the inner circulation region.


Similar results in the case of a pusher swimmer are shown in Fig.~\ref{fig_push} with flow fields displayed  on the left and their spatial decay on the right. Results are similar to the puller case except that now the toroidal vortex appears behind the swimmer. The pusher remains a pusher in the far field with $r^{-2}$ spatial decay of the velocity.

\section{\label{sec_im}Swimming under rigid confinement}
The model developed so far concerned locomotion in a two-fluid domain whose boundary was moving with the swimmer in a quasi-steady fashion. Experimentally, the study of {\it H.~pylori} locomotion demonstrated that, in the absence of the inner-viscosity fluid, the cells were in fact not able to self-propel at all  \cite{bansil09}. From a fluid mechanical point of view, this  suggests  that the viscosity in the outer fluid might be too high for the cells to move.  We might therefore propose to model the outer fluid in a simpler fashion by 
  assuming that it is, for all times, a rigid matrix with a no-slip boundary condition at the interface $r=b$. This situation is thus a special case of the calculation carried out in the previous section in the limit where the outer viscosity is much larger than the inner one.

\subsection{Squirming in a rigidly-bounded flow}
In the laboratory frame of reference, 
the boundary conditions at the swimmer surface and the interface are given for this new model by
\begin{subeqnarray}  \label{bc_im}
  v_r\vert_{r=a} &= &\sum_{n=0}^{\infty}A_n(t)P_n(\xi)+UP_1(\xi),\\
  v_{\theta}\vert_{r=a} &=& \sum_{n=1}^{\infty}B_n(t)V_n(\xi)-UV_1(\xi),\\
  v_r\vert_{r=b} &=& 0,   \\
   v_{\theta}\vert_{r=b} &=& 0.
\end{subeqnarray}

From  Lamb's general solutions (Eq.~\ref{gensol}) and 
the above boundary condition (Eq.~\ref{bc_im}),
the fluid velocity fields are solved  similarly to the work from Sec.~\ref{twofluid}, only with much simpler algebra. 
The undetermined constants in Eq.~\eqref{unconst} for $n=0, 1$
are the solutions to
\begin{subeqnarray} 
  \bar{p}_{-1} + \frac{1}{a^2}\bar{\phi}_{-1} & = & A_0,\label{s1} \\ 
  \bar{p}_{-1}+\frac{1}{b^2}\bar{\phi}_{-1} & = & 0,\label{s2}\\
  a^2\bar{p}_{1} + \bar{\phi}_{1} + \frac{1}{a}\bar{p}_{-2} +\frac{1}{a^3}\bar{\phi}_{-2} & = & A_1 + U,\label{s3}\\
  -2a^2\bar{p}_{1} - \bar{\phi}_{1} - \frac{1}{2a}\bar{p}_{-2} +\frac{1}{2a^3}\bar{\phi}_{-2} & = & B_1 - U,\label{s4}\\
  b^2\bar{p}_{1} + \bar{\phi}_{1} + \frac{1}{b}\bar{p}_{-2} +\frac{1}{b^3}\bar{\phi}_{-2} & = & 0,\label{s5}\\
  -2b^2\bar{p}_{1} - \bar{\phi}_{1} - \frac{1}{2b}\bar{p}_{-2} + \frac{1}{2b^3}\bar{\phi}_{-2} & = & 0.\label{s6}
\end{subeqnarray}
{Here the subscript $1$ is omitted since only the inner fluid exists.} 
By applying the force-free condition on the swimmer surface,
i.e. $\bar{p}_{-2}=0$, we deduce the value of the swimming velocity,
\begin{align}
  U = \frac{5\lambda^2(\lambda^3-1)}{3(\lambda^5-1)}(A_1+B_1)-2A_1-B_1,
  \label{velconf}
\end{align}
and the solutions to Eq.~\eqref{s6} are given by
\begin{subeqnarray}
  \bar{p}_{-1} & = &  -A_0\frac{1}{\lambda^2-1},\\
  \bar{\phi}_{-1} & = &  A_0\frac{a^2\lambda^2}{\lambda^2-1}, \\
  \bar{p}_{1} & = &  (A_1+B_1)\frac{1}{a^2(\lambda^5-1)}, \\
  \bar{\phi}_{1} & = &  -\frac{5}{3}(A_1+B_1)\frac{\lambda^2}{\lambda^5-1}, \\
  \bar{\phi}_{-2} & = &  \frac{2}{3}(A_1+B_1)\frac{a^3\lambda^5}{\lambda^5-1}\cdot
\end{subeqnarray}
The other  unknown constants $\bar{p}_n$ and $\bar{\phi}_n$ for $n \ge 2$ are  easily obtained as
\begin{subeqnarray}
  \bar{p}_{n} & = &  \frac{1}{a^{n+1}\Delta_n}(N_1A_n+N_2B_n),  \\
  \bar{\phi}_{n} & = &  \frac{1}{a^{n-1}\Delta_n}(N_3A_n+N_4B_n),  \\
  \bar{p}_{-(n+1)} & = &  \frac{a^n}{\Delta_n}(N_5A_n+N_6B_n),  \\
  \bar{\phi}_{-(n+1)} & = &  \frac{a^{n+2}}{\Delta_n}(N_7A_n+N_8B_n),
\end{subeqnarray}
and thus the  velocity fields are given in the laboratory frame of reference by
\begin{subeqnarray}
  v_r &=& A_0\left\{\frac{\lambda^2}{\lambda^2-1}\Big(\frac{a}{r}\Big)^2-\frac{1}{\lambda^2-1}\right\}P_0(\xi)\nonumber\\
 &&  +(A_1+B_1)\left\{\frac{1}{\lambda^5-1}\Big(\frac{r}{a}\Big)^2 -\frac{5\lambda^2}{3(\lambda^5-1)}
  +\frac{2\lambda^5}{3(\lambda^5-1)}\Big(\frac{a}{r}\Big)^3\right\}P_1(\xi) \nonumber\\
  &&+\sum_{n=2}^{\infty} \frac{1}{\Delta_n}\left\{(N_1A_n+N_2B_n)\Big(\frac{r}{a}\Big)^{n+1}
  +(N_3A_n+N_4B_n)\Big(\frac{r}{a}\Big)^{n-1}\nonumber\right.\\
  &&\qquad\left.+(N_5A_n+N_6B_n)\Big(\frac{a}{r}\Big)^{n} 
 +(N_7A_n+N_8B_n)\Big(\frac{a}{r}\Big)^{n+2}\right\} P_n(\xi), \\
  v_{\theta} &=&(A_1+B_1)\left\{-\frac{2}{\lambda^5-1}\Big(\frac{r}{a}\Big)^2+\frac{5\lambda^2}{3(\lambda^5-1)}
  +\frac{\lambda^5}{3(\lambda^5-1)}\Big(\frac{a}{r}\Big)^3\right\}V_1(\xi) \nonumber\\
  &&+\sum_{n=2}^{\infty}\frac{1}{\Delta_n}\left\{-\frac{n+3}{2}(N_1A_n+N_2B_n)\Big(\frac{r}{a}\Big)^{n+1}
  -\frac{n+1}{2}(N_3A_n+N_4B_n)\Big(\frac{r}{a}\Big)^{n-1}\right.\nonumber\\
 &&\qquad \left. +\frac{n-2}{2}(N_5A_n+N_6B_n)\Big(\frac{a}{r}\Big)^{n} 
   +\frac{n}{2}(N_7A_n+N_8B_n)\Big(\frac{a}{r}\Big)^{n+2}\right\}
  V_n(\xi),
  \label{vf_cf}
\end{subeqnarray}
where the undefined constants are listed in Appendix~\ref{table2}. We note, unsurprisingly, that both the  swimming velocity (Eq.~\ref{velconf}) and the fluid flow fields (Eq.~\ref{vf_cf}) in this rigidly confined case are  recovered from the two-fluid model (Eqs.~\ref{vel_sq} and~\ref{vf_1f} respectively) by formally taking the limit $\sigma \rightarrow \infty$

\begin{figure}[t]
  \centering
  \includegraphics[scale=0.45,angle=0]{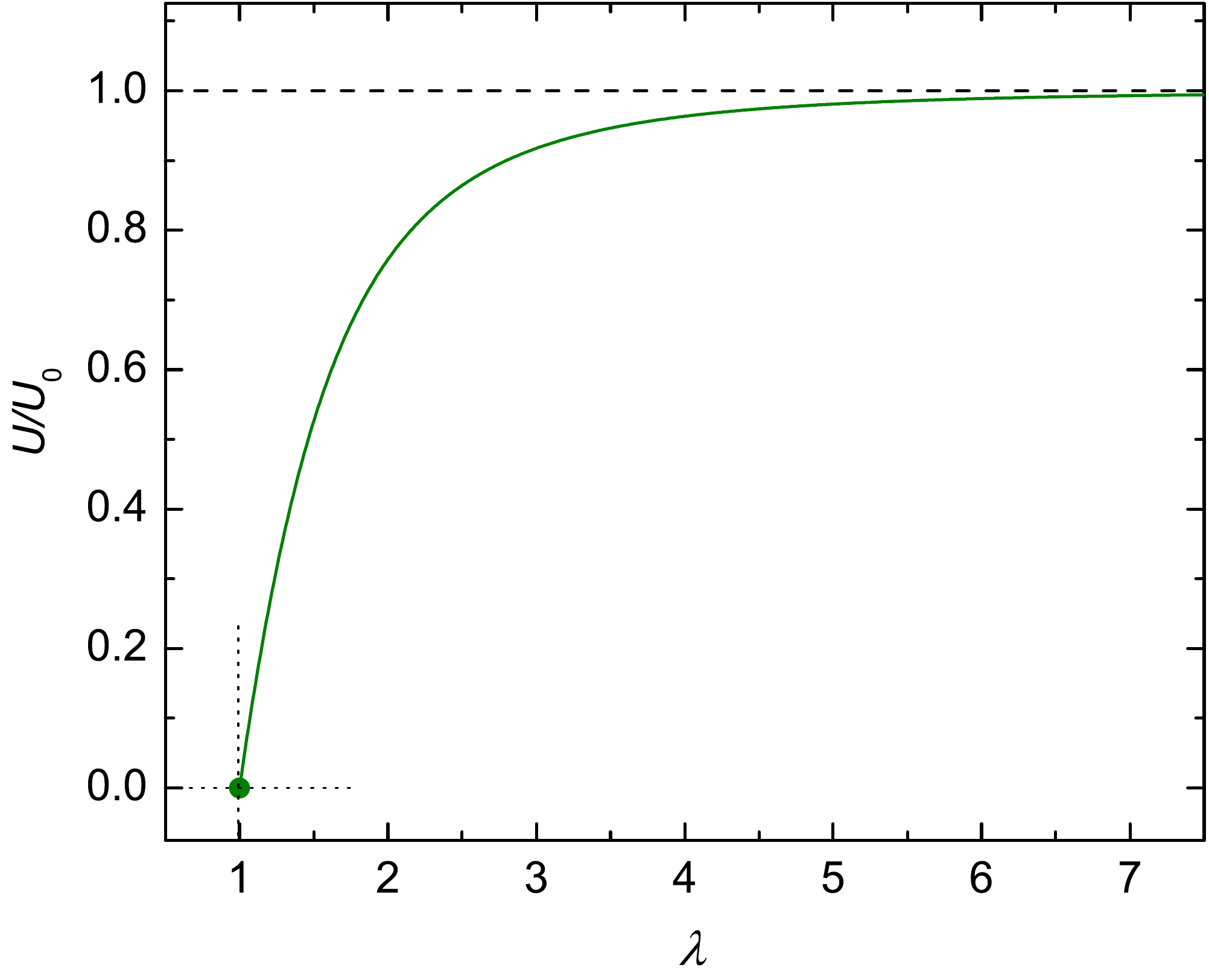}
  \caption{Swimming velocity as a function of the confinement size, $\lambda=b/a$, for squirming under rigid confinement with only tangential actuation ($A_n=0$).
    The horizontal line indicates the swimming velocity for the unbounded fluid, 
    {$U_0=2B_1/3$}.
  }
  \label{vel-conf}
\end{figure}

\begin{figure}
  \centering
  \includegraphics[scale=0.55,angle=0]{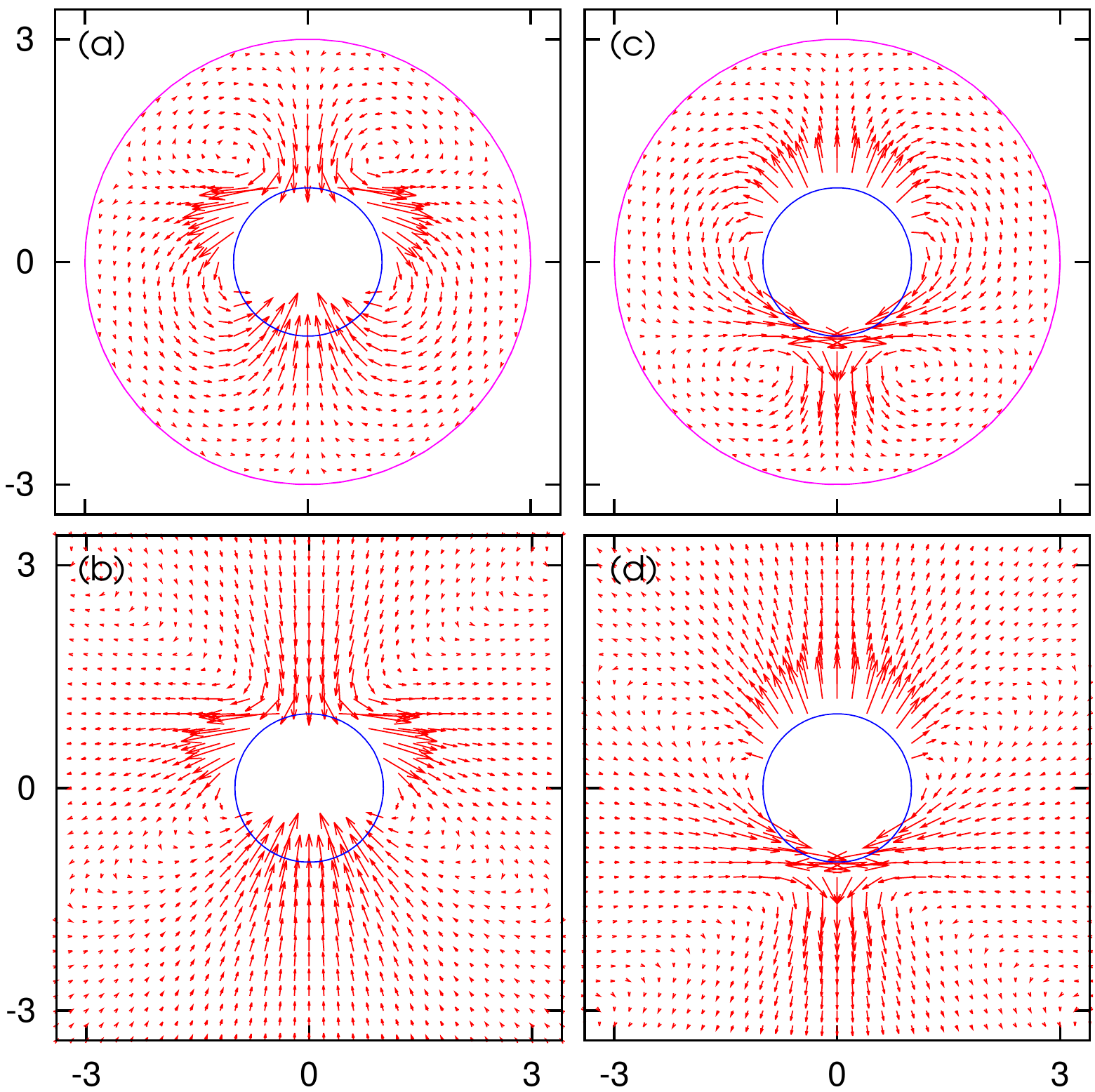}
  \caption{The fluid flow fields generated by the puller for the confinement size
    (a) $\lambda=b/a=3$ and (b) $\lambda=30$
    and by the pusher for (c) $\lambda=3$ and (d) $\lambda=30$
    in the laboratory frame of reference.
  }
  \label{mode}
\end{figure}

In Fig.~\ref{vel-conf} we plot the value of the 
swimming velocity, Eq.~\eqref{velconf},  as a function of the confinement size, $\lambda=b/a$. We consider for simplicity only the case with  tangential velocities (i.e.~$A_n=0$ for all $n$).  The swimming velocity decreases when the degree of confinement increases, a trend  consistent with {two-fluid model} (see Fig.~\ref{vel_lam}). In the case where the actuation contains only radial modes (i.e.~$B_n=0$ for all $n$), the trend is reversed and the 
the swimming speed increases with the amount of  confinement. The  flow fields generated by swimmers under rigid confinement  are shown in Fig.~\ref{mode}  for a puller (Fig.~\ref{mode}a and b)
and pusher (Fig.~\ref{mode}c and d) in both small and large  confinement. The flow patterns are comparable to those seen in the two-fluid model when the viscosity in the outer fluid is large 
(see Fig.~\ref{fig_pull} and Fig.~\ref{fig_push}) with an increase of the circulation with the degree of confinement.

\subsection{Surface deformation}
Microorganisms such as \textit{Paramecium} and \textit{Volvox} exploit the deformation of many flagella or cilia distributed on their surface and beating  in synchrony in order to self-propel~\cite{brennen77,shay99,goldstein:15}. Beyond the squirmer model, a more precise theoretical description of such types of swimmers  requires including effective  tangential and radial boundary conditions such that they both  result from the time-varying deformation of material elements on the surface of the swimmer \cite{lighthill52,blake71,stone96,ishikawa06,pak:14,pedley:16}. 
In this   section we show how to characterise this type of swimming under rigid confinement.

Assume that material points on the surface of the swimmer  oscillate with  small amplitude $\epsilon$, so that the deformed radial and tangential positions of material points at Lagrangian positions $R=a$ and $\theta=\theta_0$
may be written without loss of generality as
\begin{align}
  R &= a \bigg(1+\epsilon\sum_{n\geq 0}^\infty \alpha_n(t)P_n(\xi) \bigg), \nonumber\\
  \theta &= \theta_0 + \epsilon\sum_{n\geq 1}^\infty \beta_n(t)V_n(\xi), 
  \label{rdef}
\end{align}
where $\alpha(t)$ and $\beta(t)$ are time-varying functions~\cite{lighthill52,blake71}.

In the frame moving with the swimmer, 
the Eulerian boundary conditions for the fluid velocity at the swimmer surface
are given by the time derivatives of Eq.~\eqref{rdef} as
\begin{align}
  &\bar{v}_r(R,\theta) = a\epsilon \sum_{n\geq 0}^\infty \alpha_n^\prime P_n(\xi), \nonumber\\
  &\bar{v}_{\theta}(R,\theta) = a\epsilon \Bigg(\sum_{n\geq 1}^\infty \beta_n^\prime V_n(\xi)\Bigg)
  \Bigg(1+\epsilon\sum_{n\geq 0}^\infty \alpha_n P_n(\xi) \Bigg),
  \label{rdef2}
\end{align}
where $\bar{v}_r$ and $\bar{v}_\theta$ are the radial and tangential velocity components
 in the moving frame, {which are related to the lab-frame velocities as $\bar{v}_r=v_r-UP_1$ and $\bar{v}_\theta=v_\theta+UV_1$,}
and $\alpha^\prime$ and $\beta^\prime$ are  time derivatives for $\alpha$ and $\beta$ respectively.

Since the first-order solution at order $\epsilon$ does not provide any propulsion by symmetry~\cite{taylor51,lighthill52,blake71,Childress1981},
we need to expand 
up to the second order in $\epsilon$ using  Taylor expansion  from the undeformed state $(a,\theta_0)$ {as
\begin{align}
  \bar{\bm{v}}({R},\theta) = \bar{\bm{v}}(a,\theta_0) + (R-a)\frac{\partial \bar{\bm{v}}}{\partial r}\Big\vert_{r=a} +(\theta-\theta_0)\frac{\partial \bar{\bm{v}}}{\partial \theta}\Big\vert_{r=a} ,
\end{align}
plus higher-order terms. Here $\bar{\bm{v}}(a,\theta_0)$ is given by Eq.~\eqref{vf_cf} by replacing $A_n$ and $B_n$ by
$A_n=a\epsilon\alpha_n^\prime$ and $B_n=a\epsilon\beta_n^\prime$ respectively, 
which implies $\bar{\bm{v}}(a,\theta_0)$ is first order in $\epsilon$. 
The derivatives $\partial \bar{\bm{v}}/\partial r$ and $\partial \bar{\bm{v}}/\partial \theta$
at $r=a$, and the displacement $R-a$ and $\theta-\theta_0$ are of orders  $\epsilon$
respectively so that $(R-a)(\partial \bar{\bm{v}}/\partial r)\vert_{r=a}$ 
and $(\theta-\theta_0)(\partial \bar{\bm{v}}/\partial \theta)\vert_{r=a}$ are both at order  $\epsilon^2$. Hence
the boundary conditions at the position $(a,\theta_0)$ up to the second order} are now written by
\begin{eqnarray}
  \bar{v}_r(a,{\theta_0}) &=& a\epsilon \Bigg[ \sum_{n\ge 0}^{\infty} \alpha_n^\prime P_n
  -\frac{\partial \bar{v}_r}{\partial r}\bigg\vert_a\sum_{n\geq 0}^{\infty}\alpha_n P_n 
  -\frac{1}{a}\frac{\partial \bar{v}_r}{\partial \theta}\bigg\vert_a\sum_{n\geq 1}^{\infty}\beta_n V_n
  \Bigg], \nonumber\\
  \bar{v}_{\theta}(a,{\theta_0}) &=& a\epsilon \Bigg[ \sum_{n\ge 1}^{\infty} \beta_n^\prime V_n
  +\Bigg(\sum_{n\geq 0}^{\infty}\alpha_n P_n \Bigg)\Bigg(\epsilon\sum_{n\geq 1}^{\infty}\beta_n^\prime V_n 
   -\frac{\partial \bar{v}_{\theta}}{\partial r}\bigg\vert_a \Bigg) 
  -\frac{1}{a}\frac{\partial \bar{v}_{\theta}}{\partial \theta}\bigg\vert_a\sum_{n\geq 1}^{\infty}\beta_n V_n
  \Bigg].
  \label{bc_exp}
\end{eqnarray}

Inserting the derivatives {of the velocity, 
  $\partial \bar{\bm{v}}/\partial r\vert_{r=a}$ and 
  $\partial \bar{\bm{v}}/\partial \theta\vert_{r=a}$,  using Eq.~\eqref{vf_cf}} into Eq.~\eqref{bc_exp} leads to explicit expressions for the boundary conditions as
\begin{widetext}
\begin{align}
  &\bar{v}_r(a,{\theta_0}) = a\epsilon \Bigg[ \sum_{n\ge 0}^{\infty} \alpha_n^\prime P_n
  + 2\epsilon\Bigg(\sum_{n\geq 0}^{\infty}\alpha_n P_n\Bigg)
  \Bigg(\frac{\lambda^2}{\lambda^2-1}\alpha_0^{\prime}P_0
  +\sum_{n\geq 1}^{\infty}(\alpha_n^\prime+\beta_n^\prime )P_n \Bigg)\\
&\hspace{60pt}  -\epsilon\Bigg(\sum_{n\geq 1}^{\infty}\alpha_n^\prime
  \frac{\partial P_n}{\partial \theta} \Bigg)\Bigg(\sum_{n\geq 1}^{\infty}\beta_n V_n\Bigg)
  \Bigg], \nonumber\\
  &\bar{v}_{\theta}(a,{\theta_0}) = a\epsilon \Bigg[ \sum_{n\ge 1}^{\infty} \beta_n^\prime V_n
  +\epsilon \Bigg\{\Bigg(\sum_{n\ge 0}^{\infty}\alpha_n P_n\Bigg)
    \Bigg(\Big\{\frac{\lambda^5+4}{\lambda^5-1}(\alpha_1^\prime+\beta_1^\prime)+\beta_1^\prime\Big\}V_1
    -\sum_{n\geq 2}^{\infty}(\kappa_n\alpha_n^\prime+(\nu_n-1)\beta_n^\prime)V_n \Bigg)\nonumber\\
    &\hspace{60pt}-\Bigg(\sum_{n\geq 1}^{\infty}\beta_n V_n \Bigg)
    \Bigg(\sum_{n\geq 1}^{\infty}\beta_n^\prime \frac{\partial V_n}{\partial \theta} \Bigg)
    \Bigg\}\Bigg],
    \label{bc_rad}
\end{align}
\end{widetext}
where $\kappa_n=\chi_1(n)/\Delta_n$, $\nu_n=\chi_2(n)/\Delta_n$, and $\chi_1$ and $\chi_2$ are given in Appendix~\ref{table2}.
Rewriting these equations formally as 
\begin{align}
  \bar{v}_r(a,{\theta_0}) &= \sum_{n=0}^{\infty}\bar{A}_n(t)P_n(\xi), \nonumber\\
  \bar{v}_{\theta}(a,{\theta_0}) &= \sum_{n=1}^{\infty}\bar{B}_n(t)V_n(\xi),
  \label{bc_im2}
\end{align}
with the introduction of new coefficients $\bar{A}_n(t)$ and $\bar{B}_n(t)$,
one finds that these boundary conditions are analogous  to those with no radial deformation (Eq.~\ref{bc_im}).

Hence, the swimming velocity with radial deformation is,
up to the second order  in $\epsilon$, given by Eq.~\eqref{velconf}
by replacing ${A}_1$ and ${B}_1$ with $\bar{A}_1$ and $\bar{B}_1$ respectively
~\cite{lighthill52,blake71}.
The coefficients $\bar{A}_1$ and $\bar{B}_1$ are obtained by surface averaging
\begin{align}
  &\bar{A}_1=\frac{3}{2}\int_{-1}^{1}\bar{v}_rP_1(\xi)d\xi,\nonumber\\
  &\bar{B}_1=\frac{3}{4}\int_{-1}^{1}\bar{v}_\theta V_1(\xi)d\xi.
\end{align}
Using the properties of associate Legendre functions~\cite{lighthill52,blake71,grad07},
we obtain the detailed expressions for $\bar{A}_1$ and $\bar{B}_1$ as
\begin{widetext}
\begin{align}
  &\bar{A}_1 =  a\epsilon 
  \bigg[\alpha_1^\prime + \epsilon 
  \bigg\{ \frac{2\lambda^2}{\lambda^2-1}\alpha_0^\prime\alpha_1 + 2\alpha_0(\alpha_1^\prime+\beta_1^\prime) 
  +\sum_{n\geq 1}^\infty \frac{6}{(2n+1)(2n+3)}
  \Big\{ (n+1)\alpha_n(\alpha_{n+1}^\prime+\beta_{n+1}^\prime)  \nonumber\\
  &\hspace{50pt}+ (n+1)\alpha_{n+1}(\alpha_{n}^\prime+\beta_{n}^\prime)
  + n\alpha_n^\prime\beta_{n+1}  
  + (n+2)\alpha_{n+1}^\prime\beta_{n} \Big\} \bigg\} \bigg], \nonumber\\
  &\bar{B}_1 = a\epsilon
  \bigg[\beta_1^\prime + \epsilon 
  \bigg\{ \Big(\frac{\lambda^5+4}{\lambda^5-1}(\alpha_1^\prime+\beta_1^\prime)+\beta_1^\prime \Big)
  \Big( \alpha_0 - \frac{1}{5}\alpha_2 \Big) 
  -\frac{1}{5}\alpha_1 \big(\kappa_2\alpha_2^\prime+(\nu_2-1)\beta_2^\prime \big) \nonumber\\
  &\hspace{50pt}+\sum_{n\geq 2}^\infty \frac{3}{(2n+1)(2n+3)}
  \Big\{ \kappa_n\alpha_n^\prime\alpha_{n+1} + (\nu_n-1)\alpha_{n+1}\beta_{n}^\prime 
  -\kappa_{n+1}\alpha_n\alpha_{n+1}^\prime - (\nu_{n+1}-1)\alpha_{n}\beta_{n+1}^\prime \Big\} \nonumber\\
  &\hspace{50pt}+6\sum_{n\geq 1}^\infty \frac{(n+2)\beta_n\beta_{n+1}^\prime-n\beta_{n+1}\beta_n^\prime}{(n+1)(2n+1)(2n+3)}
  \bigg\}\bigg].
  \label{ab1}
\end{align}
\end{widetext}

One sees that these solutions reduce to those in the unbounded fluid in the limit where 
 $\lambda \gg 1$~\cite{lighthill52,blake71}.

\section{\label{bio}Biophysical modelling}

So far in our fluid mechanical model we  have assumed that the model bacterium self-propels by applying  surface velocities,   Eq.~\eqref{bc_im},  whose magnitudes were  not influenced by the presence of the confinement. In other words, we considered only the ``fixed kinematics'' limit. 
However, under confinement both biological and artificial swimmers may modify the magnitude of their actuation on the fluid. For example, flagellated bacteria are known to rotate  their helical flagella by applying a constant motor torque~\cite{berg04}. Any change to the surrounding fluid mechanics would then impact the rotation speed of the motor, hence the locomotion of the cell.

In order to model how the swimming velocity of a cell, or that of an artificial swimmer,  would respond to a self-generated confinement, three possible scenarios are now considered. In the first one, we assume that the surface velocities imposed by the swimmer remain identical for all confinement (this is the fixed-kinematics case, as above). The second situation is that of  swimming under fixed total  rate of working, which at low Reynolds number is equivalent to a fixed total dissipation in the fluid; this could be the limit under which power-control synthetic devices operate. The third scenario, directly relevant to the biological limit, is that of swimming under a fixed mean surface traction (or stress); given that the geometry of the swimmer is unchanged, this means that all local force and torque, and all of their moments, are kept constant for all confinement values, and is therefore a model for the  fixed-motor-torque actuation of flagellated bacteria \cite{berg04}.

The  rate of working of the swimmer against the fluid, or power $P$, is defined as
\begin{align}
  P = - \int\!\!\!\int_S \bm{v} \cdot \bm{\Pi} \cdot \hat{\bm{r}} d{S}  ,  
\end{align}
where 
$\bm{\Pi}$ is the stress tensor in the fluid  
and $S$ indicates the surface of the swimmer. Using our analytical calculation we can obtain an analytical expression for $P$ given by
  \begin{eqnarray}
  \frac{P}{4\pi \mu a } &=& 2\frac{2\lambda^2+1}{\lambda^2-1}A_0^2
  +\frac{2(2\lambda^5+3)}{3(\lambda^5-1)}(A_1+B_1)^2 \nonumber\\
  && +\sum_{n\ge 2}^{\infty}\frac{1}{(2n+1)\Delta_n^2(\lambda)}
  \bigg\{ 2 \Big(a_nN_{o}A_n^2+b_nN_{e}B_n^2 +(a_nN_{e}+b_nN_o)A_nB_n\Big) \nonumber\\
  &&\qquad+\frac{4}{n(n+1)}\Big(c_n\bar{N}_oA_n^2+d_n\bar{N}_{e}B_n^2 +(c_n\bar{N}_{e}+d_n\bar{N}_o)A_nB_n\Big)\bigg\}, 
\end{eqnarray}
where $\mu$ is the fluid viscosity and the other constants are given in Appendix~\ref{table2}.
Note that the expression for $P$ reduces to the Lighthill and Blake's solutions 
in the limit where $\lambda \gg 1$~\cite{lighthill52,blake71}. For a spherical squirmer with  {displacements} along the radial direction, we {may} define the mean magnitude of the   traction on the surface of the swimmer, $M$,   as 
\begin{equation}
M=\int\!\!\!\int_S \sqrt{\Pi_{rr}^2+\Pi_{r\theta}^2} dS.
\end{equation}

To be able to apply this to the biological system, we have to prescribe appropriate values for the various swimming modes. For simplicity we consider tangential swimming only and focus on the first two modes of swimming, $B_1$ and $B_2$, and take the other ones to zero. We then consider two cases for illustration. 

The first case is the one where $B_2=0$ so that there is only one surface mode. In that case, 
all expressions above may be evaluated analytically and we obtain $P = 8\pi \mu a B_1^2\kappa$ and 
 $M = 3\pi^2 \mu aB_1\kappa$, where $\kappa\equiv(2\lambda^5+3)/(3\lambda^5-3)$. The three resulting  different swimming  velocities are then given by the following expressions:
 \begin{subeqnarray}\label{table3}
        &&\text{Fixed velocity},  B_1: U/U_0 = (2\lambda^5-5\lambda^2+3)/\{2(\lambda^5-1)\},\\
        &&\text{Fixed power}, P: U/U_0 = (2\lambda^5-5\lambda^2+3)/ \sqrt{2(\lambda^5-1)(2\lambda^5+3)}, \\
        &&\text{Fixed surface stress}, M: U/U_0 = (2\lambda^5-5\lambda^2+3)/(2\lambda^5+3) .
\end{subeqnarray}

To provide a more realistic model, we introduce the second surface mode $B_2$.  By inspecting past work on the flows created by a prototypical flagellated bacterium, we can then estimate the ratio $B_2/B_1$ relevant to the biological world. Specifically, it is known that the flow created by a swimming bacterium  is that of a force-dipole, denoted $p$. For  a tangential squirmer of radius $a$ in a fluid of viscosity $\mu$, the  force dipole  is in turn known to be equal to $-4\pi \mu a^2 B_2$ \cite{ishikawa06}. Since the swimming speed is given by $U=2B_1/3$ we obtain
\begin{equation}
\frac{B_2}{B_1} =-\frac{p}{6\pi\mu U a^2} \cdot
\end{equation}
Plugging in the measured  values for a swimming {\it E.~coli} bacterium (incidently, the only prokaryote for which this measurement exists), namely  $p\approx0.8$~pN.$\mu$m \cite{drescher11}  and together with  $a\approx 1$~$\mu$m {and} $U\approx 30$~$\mu$m/s \cite{berg04} we obtain $B_2/B_1\approx -1.4$. For this second model, we then make the modelling assumption that the ratio $B_2/B_1$ remains constant for all values of the confinement ratio, $\lambda$. The expressions for the swimming velocity may then be expressed for this model as
 \begin{subeqnarray}\label{table4}
        &&\text{Fixed velocity},  B_1: U/U_0 = (2\lambda^5-5\lambda^2+3)/\{2(\lambda^5-1)\},\\
        &&\text{Fixed power}, P: U/U_0 = {[(2\lambda^5-5\lambda^2+3)/\{2(\lambda^5-1)\}](B_1^{\prime}/B_1)}, \\
        &&\text{Fixed surface stress}, M: U/U_0 = {[(2\lambda^5-5\lambda^2+3)/\{2(\lambda^5-1)\}](B_1^{\prime\prime}/B_1)}.
\end{subeqnarray}
{where $B_1^\prime$ and $B_1^{\prime\prime}$ are the modified  values of $B_1$ due to  confinement 
and where the  ratios $B_1^\prime/B_1$ and $B_1^{\prime\prime}/B_1$ can be computed from the  expressions for power and stress.}

\begin{figure}[t]
  \centering
  \includegraphics[width=0.99\textwidth]{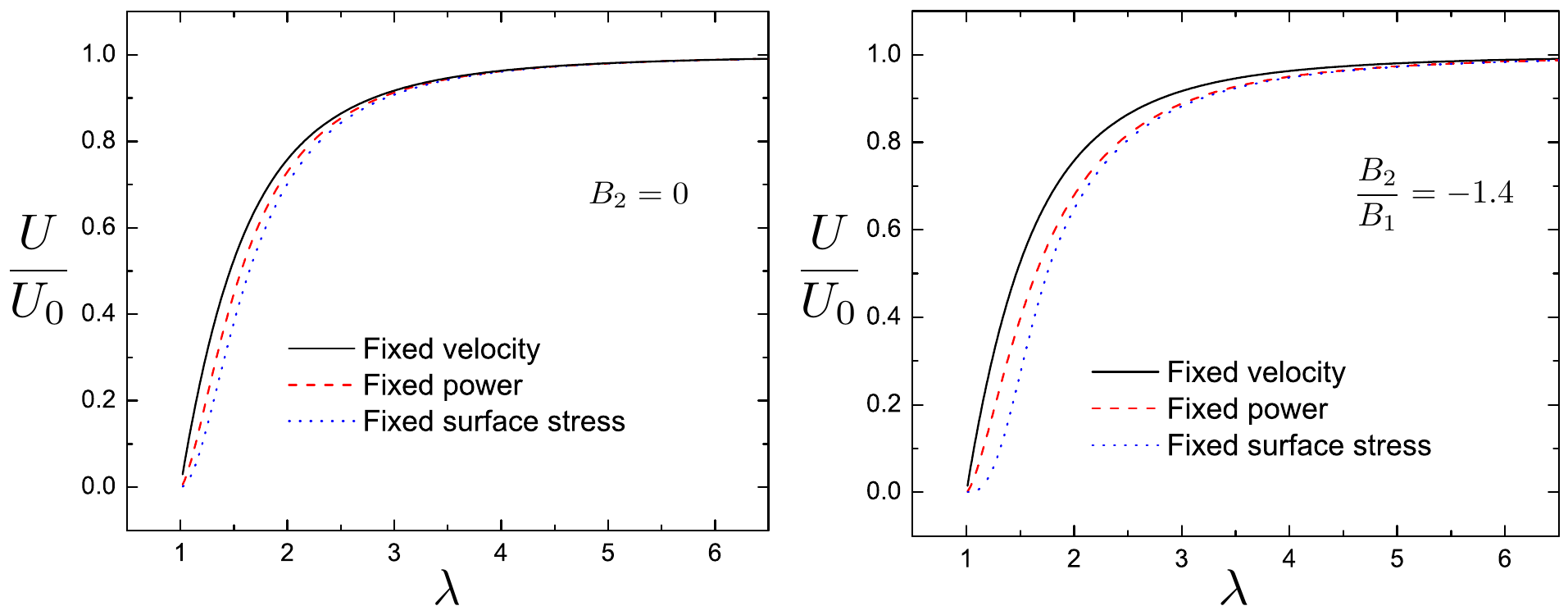}
  \caption{Swimming velocity as a function of the confinement size, $\lambda = b/a$, 
    for tangential swimming under three modelling assumptions: (a)  fixed surface velocity (solid  black lines); (b)      fixed power (dashed red lines); (c) fixed mean surface stress (dotted blue lines). Left: swimmer with no stresslet and only $n=1$ tangential mode; 
 right: swimming with two modes such that $B_2/B_1=-1.4$, appropriate value for a flagellated bacterium.
  }
  \label{vel_corr}
\end{figure}

The impact on confinement on these two models is  illustrated in Fig.~\ref{vel_corr} where we plot the swimming velocity as a function of the confinement size ratio, $\lambda = b/a$, in the three cases for the model with one mode (left) and two modes (right). In each case, we observe that the three modelling scenarios, although they are based on different physical assumptions about the method of locomotion, show very similar  dependence of the swimming speed on the degree of confinement. Furthermore, the difference between a zero and non-zero  value of $B_2$ is minimal.

\begin{table}[t]
\begin{center}
\begin{tabular}{ccccccccc}
  \hline 
Model&\quad\quad &Velocity   & \quad\quad& Range of $\lambda$   & \quad\quad& Predicted size of low-  & \quad\quad& Increase in stress\\
&&ratio  & \quad\quad& from model  & \quad\quad& viscosity region ($b$) & \quad\quad& ($M/M_0-1$)\\
  \hline 
$B_2=0$
&&$ U/U_0 \approx 0.80$  & \quad\quad& $2.16 - 2.30$  & \quad\quad& $ 9$~$\mu$m & \quad\quad&$	0.020 - 0.054$\\
&&$ U/U_0 \approx 0.90$  & \quad\quad& $2.8	- 2.92$  & \quad\quad& $ 11.4 $~$\mu$m & \quad\quad& $0.0059 - 0.015$\\
&&$ U/U_0 \approx 0.95$  & \quad\quad& $3.58	- 3.68$ & \quad\quad&$ 14.5 $~$\mu$m & \quad\quad& $(1.9 - 4.3)\times 10^{-3} $	\\
&&$ U/U_0 \approx 0.99$  & \quad\quad& $6.24	- 6.30$ & \quad\quad&$ 25$~$\mu$m & \quad\quad& $ (1.3-2.6)\times 10^{-4}$\\
&&$ U/U_0 \approx 0.999$  & \quad\quad& $11.82 - 11.86$ & \quad\quad&$ 47.4$~$\mu$m & \quad\quad& $ (0.54 - 1.1)\times10^{-5}$\\
  \hline 
$\displaystyle{B_2}/{B_1}=-1.4$
&&$ U/U_0 \approx 0.80$  & \quad\quad& $2.16 - 2.48$  & \quad\quad& $ 9.3$~$\mu$m & \quad\quad&	$0.016 - 0.13$\\
&&$ U/U_0 \approx 0.90$  & \quad\quad& $2.8 -  3.27$  & \quad\quad& $ 12.1$~$\mu$m & \quad\quad& $0.0046 - 0.049$ \\
&&$ U/U_0 \approx 0.95$  & \quad\quad& $3.58 - 4.04$ & \quad\quad&$ 15.2$~$\mu$m & \quad\quad& 	$ 0.0020 - 0.022$	\\
&&$ U/U_0 \approx 0.99$  & \quad\quad& $6.24 - 6.91$ & \quad\quad&$ 26.3$~$\mu$m & \quad\quad& $(0.34 - 4.2) \times 10^{-3}$\\
&&$ U/U_0 \approx 0.999$  & \quad\quad& $11.82 - 13.27$ & \quad\quad&$ 50.2$~$\mu$m & \quad\quad& $(0.46 - 6.2) \times 10^{-4}$\\
  \hline 
\end{tabular}
\end{center}
\caption{Predictions of our mathematical model assuming a cell size of 4~$\mu$m \cite{bansil09}. Top: Swimmer with one tangential surface mode ($B_2=0$); bottom: two modes with constant ratio $B_2/B_1=-1.4$.
\label{table:bio}
}
\end{table}%

A more detailed look is offered in Table~\ref{table:bio} where we give the range of values of the size ratio, $\lambda$, required to obtain a certain threshold of  swimming speed compared to that in an infinite fluid, $U_0$ for both models. Since the  size of a {\it H. pylori} cell  body is approximately  4~$\mu$m \cite{bansil09}, we then use our models to predict the range of sizes for the low-viscosity region, $b$, required to obtain swimming ratio at that level.  Along with this, we give in the last  column of Table~\ref{table:bio} the typical order of magnitude of the increase in swimmer stress arising from pure confinement assuming fixed kinematics {or power} (i.e.~how much more torque the cell would have to expand for this value of $\lambda$ in order to apply the same boundary conditions).

Remarkably, while the three different biophysical assumptions and the two different values for the surface modes  lead to different values in $\lambda$, the results from Table~\ref{table:bio} show that the range of values is small and all  lead to very similar predictions for the size of the low-viscosity region. In particular, the results 
 suggest that  in order for the cell to swim within 1\% of its free-swimming value (resulting in mean hydrodynamic stresses within 0.5\%), the cell  needs to liquify a spherical cloud of size above $\approx25~\mu$m. This is consistent with recent theoretical work focusing on the   physico-chemical modelling of the gel breakdown and  predicting a  low-viscosity region of   size $\approx44$~$\mu$m   \cite{mirbagheri16}. With our modelling approach, that size would lead to swimming with a velocity within 0.1\% of the speed in an unbounded fluid and a typical stress acting on the cell within 0.05\% of what would be experienced in the absence of confinement.

\section{Conclusion}
In this paper we have developed a two-fluid model for the locomotion of
\textit{H.~pylori} under self-produced confinement. The model swimmer, assumed
to be spherical, was surrounded by a shell of low viscosity fluid itself
enclosed by a  high viscosity fluid with an  interface between the two fluid
at a fixed distance from the swimmer in the co-moving frame. This model could
be solved exactly and we obtained analytical expressions for the swimming
velocity and the complete flow field. The high-viscosity outer fluid plays the
role of a soft-confining surface which decreases the value of the swimming
velocity, decreases the hydrodynamic signature in the far field, and increases
the recirculation of flow near the swimmer.    Applied to the specific case of
\textit{H.~pylori}, the model suggests that 
the swimming speed decrease resulting from the incomplete destruction of mucin
  gel in the outer region 
is generic and essentially independent
of the modelling assumptions   on the feedback of the confinement on the
swimming mode. It also suggests that in order for the bacterium to swim under
stresses similar to that experienced in unbounded fluid with no confinement, the cell needs to
generate chemically a low-viscosity region  of size at least
$\approx$~25~$\mu$m i.e.~approximately six times the  size of the 
   cell body (see Fig.~\ref{EM}). 
Beyond the case of a single cell, if multiple bacteria swim in this scenario, we expect the results in this paper to be only weakly affected by interactions provided the cells remain  separated by a distance larger than twice the size of the low-viscosity region.

Recent experiments with \textit{H.~pylori} addressed  the consequences of a variation in the chemical environment
  on cell motility    \cite{martinez2016helicobacter}, and our two-fluid model may be used to
  rationalise the observed changes in the swimming speeds.
In addition to its application to biology, the theory developed in this paper  may also explain  experimental observations for artificial magnetic micropropellers mimicking \textit{H.~pylori}'s chemical swimming strategy~\cite{walker:15}. The decrease of the swimming speed observed experimentally was attributed to the viscoelastic properties of materials
~\cite{shen11,liu:11,power:13,gagnon:13,tom:13,dasgupta:13,emily:15,yi15} but in fact our calculations show that confinement is sufficient to lead to a decrease.  Beyond magnetic actuation, a model similar to ours could be developed to study phoretic swimming under liquid confinement by combining it with the solution to a chemical transport problem. In the case of phoretic swimming, the chemical reactions involved in locomotion  might increase the chemical gradients and counter-act the increase in hydrodynamic friction under confinement~\cite{pop09}. Finally, since the model developed here is able to address finite-length swimmers, it could also be used  computationally in future work to address collective effects in complex fluids by allowing a discretisation of domains with different viscosities.


\begin{acknowledgments}
This work was funded in part by the Isaac Newton Trust, Cambridge, and by the European Union through a Marie Curie CIG grant and a ERC Consolidator grant.
\end{acknowledgments}

\appendix
\section{Coefficients for the two-fluid model}
\label{table1}
The expressions for the velocity fields in the two-fluid model, Eqs.~\ref{vf_1f}-\ref{vf_2f}, involve a series of constants, which are given by the following:

$\bullet$ $\Delta$ constants: 

\begin{subeqnarray}
  \Delta_0  &= &2(\lambda^{2}-1)\sigma+\lambda^{2}+2,  \\
  \Delta_1  &= &
  {3\{2(\lambda^5-1)\sigma+3\lambda^5+2\}},  \\
  \Delta_{n,1}  &= &(2n+1)^2
  \{(\sigma+\Omega_1)\lambda^{2n-1}-(\sigma-1)\} 
  \{(\sigma+\Omega_2)\lambda^{2n+3} -(\sigma-1)\} \nonumber\\
  &&-(2n-1)(2n+3)\{(\sigma+\Omega_1)\lambda^{2n+1}-(\sigma-1)\}^2, \\
  \Delta_{n,2} &= &(2n+1)^2(\lambda^{2n-1}-\Psi_1)(\lambda^{2n+3}-\Psi_2) 
   -(2n-1)(2n+3)(\lambda^{2n+1}-\Psi_2)^2 .
\end{subeqnarray}

$\bullet$ $\Omega$ constants:             
    \begin{subeqnarray}
        \Omega_1 &= &\di \frac{2(n-1)(n+1)}{2n^2+1}, \\
        \Omega_2 &= & \di \frac{2(n^4+2n^3-n^2-2n+3)}{n(n+2)(2n^2+1)}.
            \end{subeqnarray}
            
$\bullet$  $\Psi$ constants:                        
    \begin{subeqnarray}
        \Psi_1 &=& \di \frac{2\{n(n+2)(2n^2+1)\sigma+2(n^4+2n^3-n^2-2n+3)\}(\sigma-1)}
        {\{(2n^2+1)\sigma+2(n^2-1)\}\{2n(n+2)\sigma+2n^2+4n+3\}},\\
        \Psi_2& =& \di \frac{2n(n+2)(\sigma-1)}{2n(n+2)\sigma+2n^2+4n+3}.
            \end{subeqnarray}

  $\bullet$           $N$ constants:             
    \begin{subeqnarray}
        N_1   &=& [-(n-2)(2n+1)\{(\sigma+\Omega_1)\lambda^{2n-1}-(\sigma-1)\} \nonumber\\
        && +n(2n-1)\{(\sigma+\Omega_1)\lambda^{2n+1}-(\sigma-1)\}](\sigma-1),\\
        N_2   &=& [2(2n+1)\{(\sigma+\Omega_1)\lambda^{2n-1}-(\sigma-1)\} \nonumber\\
&& -2(2n-1)\{(\sigma+\Omega_1)\lambda^{2n+1}-(\sigma-1)\}](\sigma-1),\\
        N_3   &=& [(n-2)(2n+3)\{(\sigma+\Omega_1)\lambda^{2n+1}-(\sigma-1)\} \nonumber\\
&& -n(2n+1)\{(\sigma+\Omega_2)\lambda^{2n+3}-(\sigma-1)\}](\sigma-1),\\
        N_4   &=& [-2(2n+3)\{(\sigma+\Omega_1)\lambda^{2n+1}-(\sigma-1)\} \nonumber\\
&& +2(2n+1)\{(\sigma+\Omega_2)\lambda^{2n+3}-(\sigma-1)\}](\sigma-1),\\
        N_5   &=& [-(n+1)(2n+3)(\lambda^{2n+1}-\Psi_2)\lambda^{2n+1} 
  +(n+3)(2n+1)(\lambda^{2n+3}-\Psi_2)\lambda^{2n-1}],\\
  N_6  & = & [-2(2n+3)(\lambda^{2n+1}-\Psi_2)\lambda^{2n+1} 
  +2(2n+1)(\lambda^{2n+3}-\Psi_2)\lambda^{2n-1}],\\
        N_7  &=& [(n+1)(2n+1)(\lambda^{2n-1}-\Psi_1)\lambda^{2n+3} 
        -(n+3)(2n-1)(\lambda^{2n+1}-\Psi_2)\lambda^{2n+1}],\\
        N_8  &=& [2(2n+1)(\lambda^{2n-1}-\Psi_1)\lambda^{2n+3}
        -2(2n-1)(\lambda^{2n+1}-\Psi_2)\lambda^{2n+1}].
            \end{subeqnarray}
            
  $\bullet$           $c$ constants:             
    \begin{subeqnarray}
        c_1 & =&\di \Big(n+\frac{3}{2}\Big)\frac{N_1}{\Delta_{n,1}}\lambda^{2n+1} 
        + \Big(n+\frac{1}{2}\Big)\frac{N_3}{\Delta_{n,1}}\lambda^{2n-1} 
        + \frac{N_5}{\Delta_{n,2}}       , \\
        c_2 & =&\di \Big(n+\frac{3}{2}\Big)\frac{N_2}{\Delta_{n,1}}\lambda^{2n+1} 
        + \Big(n+\frac{1}{2}\Big)\frac{N_4}{\Delta_{n,1}}\lambda^{2n-1} 
        + \frac{N_6}{\Delta_{n,2}}    ,    \\
        c_3 & =&\di -\Big(n+\frac{1}{2}\Big)\frac{N_1}{\Delta_{n,1}}\lambda^{2n+3} 
        - \Big(n-\frac{1}{2}\Big)\frac{N_3}{\Delta_{n,1}}\lambda^{2n+1} 
        + \frac{N_7}{\Delta_{n,2}}    ,    \\
        c_4 & =&\di -\Big(n+\frac{1}{2}\Big)\frac{N_2}{\Delta_{n,1}}\lambda^{2n+3} 
        - \Big(n-\frac{1}{2}\Big)\frac{N_4}{\Delta_{n,1}}\lambda^{2n+1} 
        + \frac{N_8}{\Delta_{n,2}}        \cdot
\end{subeqnarray}

\section{Coefficients for the confined model}\label{table2}
The constants entering the solution for the velocity field in the rigidly-confined case, Eq.~\eqref{vf_cf}, are given by the following:

$\bullet$ $\Delta$ constants:

\begin{subeqnarray}
  \Delta_n = (2n+1)^2(\lambda^{2n-1}-1)(\lambda^{2n+3}-1) 
  -(2n-1)(2n+3)(\lambda^{2n+1}-1)^2.
\end{subeqnarray}

$\bullet$            $N$ constants:
    \begin{subeqnarray}
        N_1 &=& -(n-2)(2n+1)(\lambda^{2n-1}-1)+n(2n-1)(\lambda^{2n+1}-1),\\
        N_2 &=& 2(2n+1)(\lambda^{2n-1}-1) -2(2n-1)(\lambda^{2n+1}-1), \\
        N_3 &=& (n-2)(2n+3)(\lambda^{2n+1}-1) -n(2n+1)(\lambda^{2n+3}-1),\\
        N_4 &=& -2(2n+3)(\lambda^{2n+1}-1) +2(2n+1)(\lambda^{2n+3}-1), \\
        N_5 &=&  -(n+1)(2n+3)(\lambda^{2n+1}-1)\lambda^{2n+1} 
         +(n+3)(2n+1)(\lambda^{2n+3}-1) \lambda^{2n-1},\\
        N_6 &=&  -2(2n+3)(\lambda^{2n+1}-1)\lambda^{2n+1} 
        +2(2n+1)(\lambda^{2n+3}-1) \lambda^{2n-1},\\
        N_7 &=& (n+1)(2n+1)(\lambda^{2n-1}-1)\lambda^{2n+3} 
        -(n+3)(2n-1)(\lambda^{2n+1}-1) \lambda^{2n+1}, \\
        N_8 &=& 2(2n+1)(\lambda^{2n-1}-1)\lambda^{2n+3} 
        -2(2n-1)(\lambda^{2n+1}-1) \lambda^{2n+1}, \\
        N_o &=& N_1 + N_3 + N_5 + N_7, \\
        N_e &=& N_2 + N_4 + N_6 + N_8,  \\
        \bar{N}_o &=& \di-\frac{n+3}{2}N_1 - \frac{n+1}{2}N_3 + \frac{n-2}{2}N_5 + \frac{n}{2}N_7 , \\
        \bar{N}_e &=& \di-\frac{n+3}{2}N_2 - \frac{n+1}{2}N_4 + \frac{n-2}{2}N_6 + \frac{n}{2}N_8 .
    \end{subeqnarray}

    $\bullet$                 $(a_n.b_n,c_n,d_n)$ constants:
    \begin{subeqnarray}
      a_n &=& \di\bigg(\frac{2n+3}{n}-n-1 \bigg)N_1 - (n-1)N_3 
      +\bigg(n+\frac{2n-1}{n+1}\bigg)N_5 +(n+2)N_7, \\
      b_n &=&\di \bigg(\frac{2n+3}{n}-n-1 \bigg)N_2 - (n-1)N_4 
       +\bigg(n+\frac{2n-1}{n+1}\bigg)N_6  +(n+2)N_8 ,\\
      c_n &=& n(n+2)N_1 + (n-1)(n+1)N_3  
       + (n-1)(n+1)N_5 +n(n+2)N_7, \\
      d_n &=& n(n+2)N_2 + (n-1)(n+1)N_4  
       + (n-1)(n+1)N_6 + n(n+2)N_8. 
    \end{subeqnarray}

    $\bullet$ $\chi$ constants: 
    \begin{subeqnarray}
      \chi_1 &=& \frac{1}{2}[4(n+1)(n+3)+(n-2)(n+3)(2n+1)^2\lambda^{2n-1} \nonumber\\
      && -2(2n-1)(2n+1)(n^2+n-1)\lambda^{2n+1} \nonumber \\
      && +n(n+1)(2n+1)^2\lambda^{2n+3} +4n(n-2)\lambda^{4n+2} ], \\
      \chi_2 &=& 8(n+1) - 3(2n+1)^2\lambda^{2n-1} +2(2n-1)(2n+3)\lambda^{2n+1}
      \nonumber \\
      &&+(2n+1)^2\lambda^{2n+3} -8n\lambda^{4n+2}.
    \end{subeqnarray}

\bibliographystyle{unsrt}


\end{document}